\documentclass[12pt]{article}

\usepackage{amsfonts}
\usepackage{amsmath}
\usepackage{amssymb}
\usepackage{mathtools}
\usepackage{graphicx}

\def\gtwid{\mathrel{\raise.3ex\hbox{$>$\kern-.75em\lower1ex\hbox{$\sim$}}}}
\def\ltwid{\mathrel{\raise.3ex\hbox{$<$\kern-.75em\lower1ex\hbox{$\sim$}}}}
\def\square{\kern1pt\vbox{\hrule height 1.2pt\hbox{\vrule width 1.2pt\hskip 3pt
   \vbox{\vskip 6pt}\hskip 3pt\vrule width 0.6pt}\hrule height 0.6pt}\kern1pt}

\begin{document}

\begin{titlepage}

\begin{flushright}
UFIFT-QG-16-06
\end{flushright}

\vskip .5cm

\begin{center}
{\bf One loop graviton corrections to dynamical photons in de Sitter}
\end{center}

\vskip .5cm

\begin{center}
D. Glavan$^{1*}$, S. P. Miao$^{2\star}$, Tomislav Prokopec$^{3\dagger}$
and R. P. Woodard$^{4\ddagger}$
\end{center}

\vskip .5cm

\begin{center}
\it{$^{1}$ Institute of Theoretical Physics, Faculty of Physics, \\
University of Warsaw, Pasteura 5, 02-093 Warsaw, POLAND}
\end{center}

\begin{center}
\it{$^{2}$ Department of Physics, National Cheng Kung University \\
No. 1, University Road, Tainan City 70101, TAIWAN}
\end{center}

\begin{center}
\it{$^{3}$ Institute for Theoretical Physics \& Spinoza Institute \\
Center for Extreme Matter and Emergent Phenomena \\
Utrecht University, Postbus 80195, 3508 TD Utrecht \\
THE NETHERLANDS}
\end{center}

\begin{center}
\it{$^{4}$ Department of Physics, University of Florida,\\
Gainesville, FL 32611, UNITED STATES}
\end{center}

\begin{center}
ABSTRACT
\end{center}

We employ a recent, general gauge computation of the one loop
graviton contribution to the vacuum polarization on de Sitter
to solve for one loop corrections to the photon mode function.
The vacuum polarization takes the form of a gauge independent,
spin 2 contribution and a gauge dependent, spin 0 contribution.
We show that the leading secular corrections derive entirely
from the spin 2 contribution.

\begin{flushleft}
\end{flushleft}

\vskip .5cm

\begin{flushleft}
$^{*}$ e-mail: Drazen.Glavan@fuw.edu.pl \\
$^{\star}$ e-mail: spmiao5@mail.ncku.edu.tw \\
$^{\dagger}$ e-mail: T.Prokopec@uu.nl \\
$^{\ddagger}$ e-mail: woodard@phys.ufl.edu
\end{flushleft}

\end{titlepage}


\section{Introduction}
\label{intro}

Serious study of quantum field theory during inflation leaves one with a
poignant appreciation for the genius of the physicists who laid the foundations
of flat space quantum field theory during the middle of the last century. Among
other things, they settled on the S-matrix as the fundamental observable
\cite{Wheeler:1937zz,Heisenberg:1943zz}. They also showed how to carefully
define this quantity \cite{Haag:1958vt,Ruelle:1962} so that it is independent
of the choice of local field variable \cite{Borchers:1960,Borchers:1962} and
consequently, independent of the choice of gauge \cite{Kamefuchi:1961sb}.

These are powerful results whose utility can be seen in many ways. One
example is inferring quantum gravitational corrections to the Coulomb
potential of a charged particle. Naively one might find this by computing
the quantum gravitational contribution to the vacuum polarization $i[
\mbox{}^{\mu} \Pi^{\nu}](x;x')$ and then use this to quantum correct
Maxwell's equations,
\begin{equation}
\partial_{\mu} \Bigl[ \sqrt{-g} \, g^{\mu\rho} g^{\nu\sigma} F_{\rho\sigma}(x)
\Bigr] + \int \!\! d^4x' \Bigl[\mbox{}^{\mu} \Pi^{\nu}\Bigr](x;x') A_{\nu}(x')
= J^{\mu}(x) \; . \label{Maxwell}
\end{equation}
However, the vacuum polarization is highly dependent on the general coordinate
gauge in flat space background. For example, if one defines the quantum metric
as $g_{\mu\nu}(x) \equiv \eta_{\mu\nu} + \kappa h_{\mu\nu}(x)$, with $\kappa^2
\equiv 16 \pi G$, then the vacuum polarization in the 1-parameter family of
exact covariant gauges $\eta^{\rho\sigma} \partial_{\rho} h_{\sigma\nu} =
\frac{b}{2} \partial_{\nu} \eta^{\rho\sigma} h_{\rho\sigma}$ is
\cite{Leonard:2012fs,Glavan:2015ura},
\begin{equation}
i \Bigl[\mbox{}^{\mu} \Pi^{\nu}\Bigr](x;x') = \frac{\kappa^2}{384 \pi^4}
\Bigl( \frac{2b \!-\! 1}{b \!-\! 2}\Bigr)^2 \Bigl[ \eta^{\mu\nu} \partial' \!
\cdot\! \partial \!-\! {\partial'}^{\mu} \partial^{\nu}\Bigr] \partial^2
\Bigl[ \frac{ \ln(\mu^2 \Delta x^2)}{\Delta x^2} \Bigr] \; , \label{flatpi}
\end{equation}
where $\Delta x^2(x;x') \equiv \eta_{\mu\nu} (x - x')^{\mu} (x - x')^{\nu}$.
One can nonetheless derive gauge independent results for the graviton
correction to the Coulomb potential by computing the scattering amplitude for
two charged, massive particles and then solving the inverse scattering
problem to reconstruct the potential \cite{BjerrumBohr:2002sx}. The problem
for inflationary cosmology is that we presently have no analogue of the
S-matrix which has been shown to be gauge independent.

The vacuum polarization on an inflationary background --- hereafter tak\-en
to be de Sitter --- cannot be less gauge dependent than its flat space limit
(\ref{flatpi}). A possible way forward is the conjecture that there might be
no gauge dependence in the leading secular effects of solutions to the
effective field equations (\ref{Maxwell}) \cite{Miao:2012xc}. These secular
effects were first noted when one uses the simplest version of the graviton
propagator \cite{Tsamis:1992xa,Woodard:2004ut} to compute the one loop
vacuum polarization \cite{Leonard:2013xsa}. In this gauge the Coulomb potential
of a co-moving observer was found to grow with time \cite{Glavan:2013jca}.
A similar growth occurs in the electric field strength of plane wave photons
\cite{Wang:2014tza}.

This paper is the second step of checking the conjecture of secular
gauge independence. In the first step \cite{Glavan:2015ura} we computed the
one loop graviton contribution to the vacuum polarization using the graviton
propagator \cite{Mora:2012zi} in the de Sitter analogue of the same
1-parameter family of covariant, exact gauges which gave (\ref{flatpi}). In
this work our result for $i[\mbox{}^{\mu} \Pi^{\nu}](x;x')$ is used to solve
(\ref{Maxwell}) for the one loop correction to plane wave photons. Section
\ref{notation} reviews our result for the vacuum polarization and summarizes
the notation we employ. Because the graviton propagator in our gauge consists
of a transverse-traceless, spin two part and a spin zero part, it is natural
to treat each separately; section \ref{spin2} works out the spin two
contribution and section \ref{spin0} gives the spin zero contribution. Our
conclusions comprise section \ref{discuss}.


\section{Notation}
\label{notation}

The purpose of this section is to summarize notation and carry out a
preliminary general analysis. We begin by reviewing the de Sitter background,
then we reduce the effective field equation (\ref{Maxwell}) to a relation for
the one loop correction to the photon mode function. The section closes after
presenting our results \cite{Glavan:2015ura} for the structure functions.


\subsection{Background Geometry}
\label{background}

We use de Sitter open conformal coordinates with Hubble constant $H$. The
invariant element is,
\begin{equation}
ds^2 = a^2 \Bigl[ -d\eta^2 + d\vec{x} \!\cdot\! d\vec{x}\Bigr] \qquad ,
\qquad a(\eta) \equiv -\frac1{H \eta} \; .
\end{equation}
Note that the conformal time $\eta$ lies in the range $-\infty < \eta < 0$,
while each of the spatial coordinates runs from $-\infty$ to $+\infty$. We
shall many times need to refer to functions of two coordinates, $x^{\mu}$
and ${x'}^{\mu}$. In this case an unprimed scale factor is $a \equiv a(\eta)
= -\frac1{H\eta}$, while the primed scale factor is $a' \equiv a(\eta') =
-\frac1{H \eta'}$.

Our results for the structure functions depend extensively on the de Sitter
invariant bi-scalar function $y(x;x')$, whose definition is,
\begin{equation}
y(x;x') \equiv a a' H^2 \Bigl[ \Vert \vec{x} \!-\! \vec{x}'\Vert^2 - \Bigl(
\eta \!-\! \eta'\Bigr)^2 \Bigr] \equiv a a' H^2 \Delta x^2 \; .
\end{equation}
Quantum field theory propagators on de Sitter depend upon a slight modification
of $y(x;x')$ which includes an infinitesimal imaginary part to specify the
appropriate boundary conditions. The two versions we require are,
\begin{eqnarray}
y_{\scriptscriptstyle ++}(x;x') & \equiv & a a' H^2 \Bigl[ \Vert \vec{x}
\!-\! \vec{x}'\Vert^2 - \Bigl(\vert \eta \!-\! \eta'\vert \!-\! i \epsilon
\Bigr)^2 \Bigr] \; , \label{y++} \\
y_{\scriptscriptstyle +-}(x;x') & \equiv & a a' H^2 \Bigl[ \Vert \vec{x}
\!-\! \vec{x}'\Vert^2 - \Bigl(\eta \!-\! \eta' \!+\! i \epsilon \Bigr)^2
\Bigr] \; . \label{y+-}
\end{eqnarray}
Note that $y_{\scriptscriptstyle ++}(x;x')$ and $y_{\scriptscriptstyle
+-}(x;x')$ agree for $\eta < \eta'$, whereas they are complex conjugates
for $\eta > \eta'$.


\subsection{The Effective Mode Equation}
\label{effective}

It turns out that de Sitter invariance, even when it is present, complicates
rather than simplifies representations of the tensor structure of the vacuum
polarization \cite{Leonard:2012si}. We therefore employed the simple, but
noncovariant, representation which was introduced to represent the vacuum
polarization from scalar quantum electrodynamics \cite{Prokopec:2002uw},
\begin{eqnarray}
\lefteqn{i\Bigl[\mbox{}^{\mu} \Pi^{\nu}\Bigr](x;x') = \Bigl( \eta^{\mu\nu}
\eta^{\rho\sigma} \!-\! \eta^{\mu\sigma} \eta^{\nu\rho}\Bigr) \partial_{\rho}
\partial_{\sigma}' F(x;x') } \nonumber \\
& & \hspace{6cm} + \Bigl( \overline{\eta}^{\mu\nu} \overline{\eta}^{\rho\sigma}
\!-\! \overline{\eta}^{\mu\sigma} \overline{\eta}^{\nu\rho}\Bigr)
\partial_{\rho} \partial_{\sigma}' G(x;x') \; , \qquad \label{vacrep}
\end{eqnarray}
where $\overline{\eta}^{\mu\nu} \equiv \eta^{\mu\nu} + \delta^{\mu}_{~0}
\delta^{\nu}_{~0}$ is the purely spatial part of the Minkowski metric. The
transformation to a de Sitter covariant representation has been worked out
\cite{Leonard:2012ex} and could be employed if desired.

Substituting (\ref{vacrep}) and $g_{\mu\nu} = a^2 \eta_{\mu\nu}$ into the
effective Maxwell equation (\ref{Maxwell}), and then performing some partial
integrations, gives an equation in terms of the field strength tensor,
\begin{equation}
\partial_{\nu} F^{\nu\mu}(x) + \partial_{\nu} \! \int \!\! d^4x' \Biggl\{
iF(x;x') F^{\nu\mu}(x') \!+\! iG(x;x') \overline{F}^{\nu\mu}(x') \Biggr\}
= J^{\mu}(x) \; . \label{Max2}
\end{equation}
(Here and henceforth we raise and lower indices with the Minkowski metric
so $F^{\mu\nu} \equiv \eta^{\mu\rho} \eta^{\nu\sigma} F_{\rho\sigma}$ and
$\overline{F}^{\mu\nu} \equiv \overline{\eta}^{\mu\rho}
\overline{\eta}^{\nu\sigma} F_{\rho \sigma}$.) By setting $J^{\mu}(x) = 0$
we see that the $\mu = 0$ component of (\ref{Max2}) is obeyed by a solution
of the form,
\begin{equation}
A_0(x) = 0 \qquad , \qquad A_i(x) = u(\eta,k) \epsilon_i(\vec{k}) e^{i
\vec{k} \cdot \vec{x}} \qquad , \qquad k_i \epsilon_i(\vec{k}) = 0 \; .
\label{ansatz}
\end{equation}
Substituting (\ref{ansatz}) into (\ref{Max2}) and factoring out both the
polarization vector and the spatial plane wave factor gives rise to the
effective mode equation,
\begin{eqnarray}
\lefteqn{ (\partial_0^2 \!+\! k^2) u(\eta,k) = -\partial_0 \! \int \!\!
d^4x' \, iF(x;x') \partial_0' u(\eta',k) e^{-i \vec{k} \cdot \Delta \vec{x}}
} \nonumber \\
& & \hspace{3.5cm} - k^2 \! \int \!\! d^4x' \Bigl[i F(x;x') \!+\! iG(x;x')
\Bigr] u(\eta',k) e^{-i \vec{k} \cdot \Delta \vec{x}} \; , \qquad
\label{modeeqn}
\end{eqnarray}
where $\Delta \vec{x} \equiv \vec{x} \!-\! \vec{x}'$.

Relation (\ref{modeeqn}) is valid to all orders. However, the structure
functions $F(x;x)$ and $G(x;x')$ are only known at order $\kappa^2$.
We therefore expand the mode function in powers of $\kappa^2$ as,
\begin{equation}
\hspace{-0.0cm}
 u(\eta,k)\! = u_{(0)}(\eta,k)\!+\!u_{(1)}(\eta,k)  \!+\! {\cal O}(\kappa^4)
,\; \big(u_{(0)}(\eta,k)\propto \kappa^0,\; u_{(1)}(\eta,k) \propto \kappa^2\big)\!\!\!\!\!\!
\label{u(eta,k) seggregation}
\end{equation}
and segregate to first order,
\begin{eqnarray}
\lefteqn{ (\partial_0^2 \!+\! k^2) u_{(1)}(\eta,k) = -\partial_0 \! \int \!\!
d^4x' \, iF^{(1)}(x;x') \partial_0' u_{(0)}(\eta',k) e^{-i \vec{k} \cdot \Delta
\vec{x}}} \nonumber \\
& & \hspace{2.5cm} - {k^2} \! \int \!\! d^4x' \Bigl[i F^{(1)}(x;x') \!+\!
iG^{(1)}(x;x') \Bigr] u_0(\eta',k) e^{-i \vec{k} \cdot \Delta \vec{x}}
\; ,
\qquad
\label{oneloop}
\end{eqnarray}
where the tree order mode function is the usual plane wave,
\begin{equation}
   u_{(0)}(\eta,k) = \frac{e^{-ik \eta}}{\sqrt{2 k}}
\,.
\label{mode function tree level}
\end{equation}
The sort of secular correction we seek is $u_{(1)}(\eta,k) \sim \ln(a)/a$, which
means the right hand side of (\ref{oneloop}) must grow like $a$. Any slower
growth does not contribute to the leading secular effect.


\subsection{Structure Functions}
\label{struct}

In an earlier work \cite{Glavan:2015ura} we applied a general gauge propagator
\cite{Mora:2012zi} to evaluate the one loop graviton contribution to the vacuum
polarization. The computation was made with Einstein + Maxwell using dimensional
regularization. Of course Einstein + Maxwell is not perturbatively renormalizable
\cite{Deser:1974zzd,Deser:1974cz} but its divergences can still be absorbed into
local higher derivative counterterms, according to the technique of Bogoliubov,
Parasiuk \cite{Bogoliubov:1957gp}, Hepp \cite{Hepp:1966eg} and Zimmermann
\cite{Zimmermann:1968mu,Zimmermann:1969jj}. Our one loop computation required
three such counterterms and their finite parts can be regarded as parameterizing
our ignorance of the ultraviolet completion of gravity + electromagnetism in the
standard sense of effective field theory \cite{Donoghue:1993eb,Donoghue:1994dn}.
Reliable results are still derivable at late times because the counterterms show
no secular increase. Focussing on the late time regime is also necessary because
we have not perturbatively correctioned the initial state from free vacuum
\cite{Kahya:2009sz}.


Our graviton propagator consists of a transverse-traceless, spin two term and
a spin zero term on which all the gauge dependence resides \cite{Mora:2012zi}.
Only a single graviton propagator enters the vacuum polarization at one loop
so it makes sense to report results for the spin two and spin zero contributions
separately. The spin two contribution to $F(x;x')$ was found to be,
\begin{eqnarray}
\lefteqn{ F^{(1)}_2(x;x') = \frac{85 \kappa^2 H^2}{72 \pi^2} \, \ln(a)
i\delta^4(x \!-\! x') - \frac{\kappa^2 H^2}{ 16 \pi^4} \Bigl[
\ln\Bigl(\frac{a a'}4\Bigr) \!+\! \frac13 \!-\! 2\gamma\Bigr] \,
\nabla^2 \Bigl( \frac1{\Delta x^2} \Bigr) } \nonumber \\
& & \hspace{-.5cm} + \frac{5 \kappa^2 H^2}{144 \pi^4} \, \partial^2
\Bigl( \frac{ \ln(\mu^2 \Delta x^2)}{\Delta x^2} \Bigr)
- \frac{5 \kappa^2 H^6 (a a')^2}{144 \pi^4} \Biggl\{
\frac{\mathcal{L}(y)}{2} \!+\! \frac{2 (2 \!-\! y) \ln(\frac{y}4)}{
4 y \!-\! y^2} \!+\! \frac{2}{y} \Biggr\} , \qquad \label{F2ren}
\end{eqnarray}
where we define the function $\mathcal{L}(y)$ as,
\begin{equation}
\mathcal{L}(y) \equiv {\rm Li}_2\Bigl(\frac{y}4\Bigr) +
\ln\Bigl(1 \!-\! \frac{y}4\Bigr) \ln\Bigl(\frac{y}4\Bigr)
- \frac12 \ln^2\Bigl(\frac{y}4\Bigr) \; . \label{Ldef}
\end{equation}
Here ${\rm Li}_{2}(z)$ is the dilogarithm function,
\begin{equation}
{\rm Li}_2(z) \equiv -\int_0^z \!\!\! dt \,
\frac{\ln(1\!-\!t)}{t} = \sum_{n=1}^{\infty} \frac{z^n}{n^2} \; .
\label{dilog}
\end{equation}
The spin two contribution to $G(x;x')$ is,
\begin{eqnarray}
\lefteqn{ G^{(1)}_2(x;x') = -\frac{5 \kappa^2 H^2}{4 \pi^2} \, \ln(a)
i\delta^4(x \!-\! x') + \frac{\kappa^2 H^2}{ 24 \pi^4} \Bigl[
\ln\Bigl(\frac{a a'}4\Bigr) \!+\! \frac13 \!-\! 2 \gamma\Bigr] \,
\nabla^2 \Bigl( \frac1{\Delta x^2} \Bigr) } \nonumber \\
& & \hspace{-.7cm} + \frac{\kappa^2 H^4 a a'}{96 \pi^4} (\partial^2_0
\!\!+\!\! \nabla^2) \!\ln(\!H^2\! \Delta x^2\!) \!+\! \frac{5
\kappa^2 H^6 (a a')^2}{72 \pi^4} \Biggl\{\!\! \frac{(1 \!-\! y)
\mathcal{L}(y)}{4} \!+\! \frac{(y \!-\! 3) \ln(\frac{y}4)}{4 \!-\! y}
\! \Biggr\}\! . \qquad \label{G2ren}
\end{eqnarray}


Next, the spin zero contributions to $F(x;x')$ and $G(x;x')$ are~\footnote{The result~(\ref{F0++})
differs slightly from formula~(202) of Ref.~\cite{Glavan:2015ura} (the second term on the first line of Eq.~(202)
ought to be multiplied by $[-\ln(a)]$ and the sign of the first term on the third line of Eq.~(202)
of Ref.~\cite{Glavan:2015ura} ought to be switched).
The result for $G_0^{(1)}$ given in~(\ref{G0++}) agrees with Eq.~(203) of Ref.~\cite{Glavan:2015ura}.},
\begin{eqnarray}
F_0^{(1)}(x;x') &= &
	 \frac{\beta^2 \kappa^2H^2}{4}
	\Biggl\{ \frac{\ln(a)}{48\pi^2 aa'} \frac{\partial^2}{H^2} i\delta^4(x\!-\!x')
	- \frac{(\beta\!-\!5)}{72\pi^2} \ln(a) i\delta^4(x\!-\!x')
\nonumber \\
& & \hspace{1.6cm}
	- \frac{1}{48\pi^2a} \frac{\partial_0}{H} i\delta^4(x\!-\!x')
	+ \frac{1}{384\pi^4} \frac{\partial^4}{aa'} \biggl[
		\frac{\ln \bigl( \frac{H^2}{4}\Delta x^2 \bigr)}
			{H^2\Delta x^2} \biggr]
\nonumber \\
& & \hspace{1.6cm}
	+ \frac{(\beta\!-\!5)}{576\pi^4} \partial^2 \biggl[
		\frac{\ln \bigl( \frac{H^2}{4}\Delta x^2 \bigr)}{\Delta x^2} \biggr]
	\!-\! \frac{H^4(aa')^2}{6\pi^4} \mathcal{N}_{F}(y)
	\Biggr\}
 ,\quad
\label{F0++}\\
G_0^{(1)}(x;x')
	&=& \frac{\beta^2 \kappa^2H^2}{4}
	\Biggl\{ \frac{[1 \!-\! \ln(a)]}{24\pi^2} i\delta^4(x\!-\!x')
	- \frac{\partial^2}{192\pi^4} \biggl[
		\frac{\ln \bigl( \frac{H^2}{4}\Delta x^2 \bigr)}{\Delta x^2} \biggr]
\nonumber \\
& & \hspace{1.6cm}
	+ \frac{H^4(aa')^2}{12\pi^4} \mathcal{N}_{G}(y)
	\Biggr\} \, , \qquad
\label{G0++}
\end{eqnarray}
Here $\mathcal{N}_F(y)$ and $\mathcal{N}_G(y)$ are complicated functions which can be represented by following series,
\begin{eqnarray}
\mathcal{N}_{F}(y) & = &
	\frac{\partial}{\partial\beta} \biggl[
	- \frac{q_0A_0}{y} \ln\bigl( \tfrac{y}{4} \bigr)
	- \frac{q_0(A_0\!+\!B_0)}{y}
	+ \frac{q_1A_1}{2} \ln^2 \bigl( \tfrac{y}{4} \bigr)
	+ q_1B_1 \ln \bigl( \tfrac{y}{4} \bigr)
\nonumber \\
& & \hspace{0.8cm}
	+ S_{F}(y) \ln \bigl( \tfrac{y}{4} \bigr)
	+ \widetilde{S}_{F}(y) \biggr] \, ,
\label{series NFy}
\\
\mathcal{N}_{G}(y) & = &
	\frac{\partial}{\partial\beta} \biggl[
	- \frac{q_0A_0}{2} \ln^2\bigl( \tfrac{y}{4} \bigr)
	- q_0(2A_0 \!+\! B_0) \ln \bigl( \tfrac{y}{4} \bigr)
\nonumber \\
& & \hspace{0.8cm}
	+ q_1A_1 y \ln^2 \bigl( \tfrac{y}{4} \bigr)
	+ q_1 (2B_1 \!-\! A_1) y \ln \bigl( \tfrac{y}{4} \bigr)
\nonumber \\
& & \hspace{0.8cm}
	+ q_1(A_1 \!-\! B_1) y
	+ S_{G}(y) \ln \bigl( \tfrac{y}{4} \bigr)
	+ \widetilde{S}_{G}(y) \biggr] \, ,
\label{series NGy}
\end{eqnarray}
where the four power series are,
\begin{eqnarray}
S_{F}(y) & = &
	\sum_{n=0}^{\infty} \frac{q_{n+2} A_{n+2}}{(n\!+\!1)} y^{n+1} \, ,
\label{SF}
\\
S_{G}(y) & = &
	\sum_{n=0}^{\infty}
	\frac{(n\!+\!3)q_{n+2} A_{n+2}}{(n\!+\!1)(n\!+\!2)} y^{n+2} \, ,
\label{SG}
\\
\widetilde{S}_{F}(y) & = &
	\sum_{n=0}^{\infty} \frac{q_{n+2}}{(n\!+\!1)}
	\biggl[ B_{n+2} - \frac{A_{n+2}}{(n\!+\!1)} \biggr] y^{n+1} \, ,
\label{StildeF}
\\
\widetilde{S}_{G}(y) & = &
	\sum_{n=0}^{\infty}
	\frac{(n\!+\!3)q_{n+2}}{(n\!+\!1)(n\!+\!2)}
	\biggl[ B_{n+2}
		\!-\! \frac{n^2 \!+\! 6n \!+\! 7}{(n\!+\!1)(n\!+\!2)(n\!+\!3)}A_{n+2} \biggr]
	y^{n+2} \! ,
\label{StildeG}
\end{eqnarray}
and the coefficients are,
\begin{eqnarray}
q_n & = & \frac{\Gamma \bigl( \frac{5}{2} \!+\! b_{N} \!+\! n \bigr)
		\Gamma \bigl( \frac{5}{2} \!-\! b_{N} \!+\! n \bigr)}
	{4^{n+1}(n\!+\!1)! \, (n\!+\!2)! \, \Gamma\bigl( \frac{1}{2} \!+\! b_N \bigr)
		\Gamma\bigl( \frac{1}{2} \!-\! b_N \bigr)} \, ,
\label{q_n}
\\
A_n & = & \frac{(n\!+\!1)}{8(n\!+\!3)(n\!+\!4)\beta}
	\Bigl[ n(n\!-\!1)\beta^2 - 4(n\!-\!1)(3n\!+\!2)\beta + 40n(n\!+\!1) \Bigr] \, ,
\label{A_n}
\\
B_n & = & A_n \Bigl[ \psi\Bigl( \tfrac{5}{2} \!+\! b_N \!+\! n \Bigr)
	+ \psi\Bigl( \tfrac{5}{2} \!-\! b_N \!+\! n \Bigr)
	- \psi(n\!+\!2) - \psi(n\!+\!3) \Bigr]
\nonumber \\
&  & + \frac{1}{8(n\!+\!3)^2(n\!+\!4)^2\beta} \Bigl[
	\beta^2(n^4 \!+\! 14n^3 \!+\! 37n^2 \!-\! 12)
\label{B_n} \\
& &	\hspace{0.cm}
	-4\beta (3n^4 \!+\! 42n^3 \!+\! 125n^2 \!+\! 52n \!-\! 22)
	\!+\! 40(n\!+\!1) (n^3 \!+\! 13n^2 \!+\! 36n \!+\! 12) \Bigr]
 .
\nonumber
\end{eqnarray}
In order to perform the computation here we need to resum the series
(\ref{SF}--\ref{StildeG}). For the purpose of this paper, in which we need the retarded
vacuum polarization, it suffices to sum only the series ${S}_{F}(y)$ and $S_G(y)$
 which multiply $\log(y)$ in Eqs.~(\ref{series NFy}) and~(\ref{series NGy}). The results can be expressed in terms of
generalized hypergeometric functions,
\begin{eqnarray}
S_F(y) & = & - \frac{(\beta\!-\!4)(\beta\!-\!6)(\beta^2\!-\!20\beta\!+\!40)}
		{128\times 5!}
\nonumber \\
& & + \frac{(\beta\!-\!4)(\beta\!-\!6)(\beta^2\!-\!12\beta\!+\!40)}
		{128\times 5!} \, {}_2F_1
	\Bigl( \bigl\{ \tfrac{7}{2} \!+\! b_N, \tfrac{7}{2} \!-\! b_N \bigr\} ,
	\bigl\{ 6 \bigr\}, \tfrac{y}{4} \Bigr)
\nonumber \\
& & - \frac{\beta(\beta\!-\!4)(\beta\!-\!6)}
		{16\times 5!} \, {}_3F_2
	\Bigl( \bigl\{ \tfrac{7}{2} \!+\! b_N, \tfrac{7}{2} \!-\! b_N, 1 \bigr\} ,
	\bigl\{ 6, 2 \bigr\}, \tfrac{y}{4} \Bigr)
\label{S_F} \\
& & + \frac{5(\beta\!+\!6)(\beta\!-\!4)(\beta\!-\!6)}
		{32\times 6!} \, y \times {}_4F_3
	\Bigl( \bigl\{ \tfrac{9}{2} \!+\! b_N, \tfrac{9}{2} \!-\! b_N, 1, 1 \bigr\} ,
	\bigl\{ 7, 2, 2 \bigr\}, \tfrac{y}{4} \Bigr) \, ,
\nonumber
\end{eqnarray}
\begin{eqnarray}
S_G(y) & = & - \frac{\beta(\beta\!-\!4)(\beta\!-\!6)(\beta\!-\!20)}{64\!\times\!5!}\, y
\nonumber \\
&+ &\!\!\!	 \frac{(\beta\!-\!4)(\beta\!-\!6)(\beta^2\!-\!12\beta\!+\!40)}{128\!\times\! 5!} \,
	y \times
	{}_{2}F_1 \Bigl( \bigl\{ \tfrac{7}{2}\!+\! b_N, \tfrac{7}{2}\!-\!b_N \bigr\},
		\bigl\{ 6 \bigr\}, \tfrac{y}{4} \Bigr)
\nonumber \\
&+ &\!\!\!\!	 \frac{(\beta\!-\!4)(\beta\!-\!6)(\beta^2\!-\!20\beta\!-\!40)}{128\!\times\! 5!} \,
	y \!\times\!
	{}_{3}F_2 \Bigl( \bigl\{ \tfrac{7}{2}\!+\! b_N, \tfrac{7}{2}\!-\!b_N, 1 \bigr\},
		\bigl\{ 6, 2 \bigr\}, \tfrac{y}{4} \Bigr)
\nonumber \\
&- &	\!\!\! \frac{\beta(\beta\!-\!4)(\beta\!-\!6)}{16\!\times\!5!} \, y\times
	{}_4F_3 \Bigl( \bigl\{ \tfrac{7}{2}\!+\! b_N, \tfrac{7}{2}\!-\!b_N, 1, 1 \bigr\},
		\bigl\{ 6, 2, 2 \bigr\}, \tfrac{y}{4} \Bigr)
\label{S_G} \\
&+ &	\!\!\! \frac{5(\beta\!+\!6)(\beta\!-\!4)(\beta\!-\!6)}{16\!\times\! 6!} \,
	y^2 \!\times\!
	{}_{4}F_3 \Bigl( \bigl\{ \tfrac{9}{2}\!+\! b_N, \tfrac{9}{2}\!-\!b_N, 1, 1 \bigr\},
		\bigl\{ 7, 2, 2 \bigr\}, \tfrac{y}{4} \Bigr)
,
\nonumber
\end{eqnarray}
where
\begin{equation}
\beta = 2\frac{2b-1}{b-2}\,,\qquad
     b_N = \sqrt{\frac{25}{4}-\beta}=\sqrt{\frac{3(3b-14)}{4(b-2)}}
\,.
\label{bN}
\end{equation}
The third line in Eq.~(\ref{S_G}) can also be written as,
\begin{equation}
y\! \times\! {}_{3}F_2
	\Bigl( \bigl\{ \tfrac{7}{2}\!+\! b_N, \tfrac{7}{2}\!-\!b_N, 1 \bigr\},
		\bigl\{ 6, 2 \bigr\}, \tfrac{y}{4} \Bigr)
= \frac{20}{\beta} \Bigl[
	{}_2F_1\Bigl( \bigl\{ \tfrac{5}{2}\!+\!b_N, \tfrac{5}{2}\!-\! b_N \bigr\},
		\bigl\{ 5 \bigr\}, \tfrac{y}{4} \Bigr)\! -\! 1 \, \Bigr]. \;
\nonumber
\end{equation}
In the rest of this work we shall use these expressions to analyse the one-loop correction to
the photon wave function~(\ref{oneloop})
arising from the one-loop graviton fluctuations on de Sitter space.
In section \ref{spin2} we discuss the spin two contribution and in section \ref{spin0}
 the spin zero contribution.


\section{Spin Two Contribution}
\label{spin2}

The purpose of this section is to work out the leading secular contribution
to the source integrals on the right hand side of (\ref{oneloop}) from the
spin two structure functions. We begin by converting the in-out structure
functions, (\ref{F2ren}) and (\ref{G2ren}), to Schwinger-Keldysh form. This
leads to Table~\ref{SKterms} of seven temporal and eight spatial terms. The
next step is substituting each term into the effective mode equation
(\ref{oneloop}) and performing the angular integrations. The total
contribution from terms 1-3 are obvious at this stage, however, some
analysis is required before the leading secular contribution can be extracted
from terms 4-7 and 8.

\subsection{Schwinger-Keldysh Structure Functions}

\begin{table}
\centering
\begin{tabular}{|c|c|c|}
\hline
$k$ & Terms from $-i F(x;x')$ & Terms from $-i [F(x;x') \!+\! G(x;x')]$ \\
\hline
1 & $85 \pi \ln(a) \delta^4(x \!-\! x')$ & $-5 \pi \ln(a)
\delta^4(x \!-\! x')$ \\
\hline
2 & $\!\!\!\!9 [\ln(\frac14 a a') \!+\! \frac13 \!-\! 2 \gamma] \nabla^2
[\frac1{2r} \delta(\Delta \eta \!-\! r)] \!\!\!\!$ & $\!\!\!\!3 [\ln(\frac14 a a')
\!+\! \frac13 \!-\! 2 \gamma] \nabla^2 [\frac1{2r} \delta( \Delta \eta \!-\! r)]
\!\!\!\!$ \\
\hline
3 & $\frac54 \partial^4 \{ \Theta [\ln[H^2 (\Delta \eta^2 \!-\! r^2)] \!-\! 1]\}$
& $\frac54 \partial^4 \{ \Theta [\ln[H^2 (\Delta \eta^2 \!-\! r^2)] \!-\! 1]\}$ \\
\hline
4 & $-\frac52 (H^2 a a')^2 \Theta \ln[\frac{(\eta + \eta')^2 - r^2}{\Delta \eta^2
- r^2}]$ & $\!\!\!\! \frac52 (H^2 a a')^2 \Theta (\Delta \eta^2 \!-\! r^2)
\ln[\frac{(\eta + \eta')^2 - r^2}{\Delta \eta^2 - r^2}] \!\!\!\!$ \\
\hline
5 & $5 H^2 a a' [\ln(\frac14 a a') \!+\! 2] \frac1{2r} \delta(\Delta \eta
\!-\! r)$ & $5 H^2 a a' [\ln(\frac14 a a') \!+\! 2] \frac1{2r} \delta(\Delta
\eta \!-\! r)$ \\
\hline
6 & $\frac{5 H^2 a a' \Theta}{(\eta + \eta')^2 - r^2}$ &
$\frac{15 H^2 a a' \Theta}{(\eta + \eta')^2 - r^2} - 10 (H^2 a a')^2 \Theta$ \\
\hline
7 & $\!\!\!\!-\frac54 H^2 a a' \partial^2 \{ \Theta [\ln[H^2 (\Delta \eta^2
\!-\! r^2)] \!-\! 1]\} \!\!\!\!$ & $\!\!\!\!-\frac54 H^2 a a' \partial^2 \{
\Theta [\ln[H^2 (\Delta \eta^2 \!-\! r^2)] \!-\! 1]\} \!\!\!\!$ \\
\hline
8 & $0$ & $\frac32 H^2 a a' (\partial_0^2 \!+\! \nabla^2) \Theta$ \\
\hline
\end{tabular}
\caption{Different terms in the temporal and spatial parts of the
Schwinger-Keldysh structure functions. To save space we have defined
$\Theta \equiv \theta(\Delta \eta \!-\! r)$, and extracted a common factor
of $\frac{\kappa^2 H^2}{72 \pi^3}$ from each term. \label{SKterms}}
\end{table}

We employ the Schwinger-Keldysh formalism \cite{Schwinger:1960qe,
Mahanthappa:1962ex,Bakshi:1962dv,Bakshi:1963bn,Keldysh:1964ud} to obtain
effective field equations which are both real and causal \cite{Chou:1984es,
Jordan:1986ug,Calzetta:1986ey}. Expressions (\ref{F2ren}) and (\ref{G2ren})
give the in-out structure functions. The procedure for converting them to
Schwinger-Keldysh form is simple \cite{Ford:2004wc}:
\begin{itemize}
\item{Derive the $++$ structure functions by replacing each factor of the
de Sitter length function $y(x;x')$ by $y_{\scriptscriptstyle ++}(x;x)$
as defined in expression (\ref{y++});}
\item{Derive the $+-$ structure function by dropping the delta function
terms, adding an overall minus sign, and replacing $y(x;x')$ by
$y_{\scriptscriptstyle +-}(x;x')$ as defined in expression (\ref{y+-}); and}
\item{Adding the $++$ and $+-$ structure functions.}
\end{itemize}
When $\eta < \eta'$ the $y_{\scriptscriptstyle ++}(x;x')$ and $y_{
\scriptscriptstyle +-}(x;x')$ agree so the $++$ and $+-$ structure functions
cancel. For infinitesimal $\epsilon$ they also cancel whenever $r \equiv
\Vert \vec{x} - \vec{x}'\Vert > \Delta \eta \equiv \eta - \eta'$. Hence
the Schwinger-Keldysh structure function vanishes unless the point
${x'}^{\mu}$ lies on or within the past light-cone of $x^{\mu}$. Because
$y_{\scriptscriptstyle ++}(x;x')$ and $y_{\scriptscriptstyle +-}(x;x')$ are
complex conjugates in this region, the sum of $i$ times the two structure
functions is real.

Our results for the Schwinger-Keldshy structure functions are reported in
Table~\ref{SKterms}. As an example, consider the contribution to
$-iF^{(1)}_{2}(x;x')$ from the prepenultimate term of expression (\ref{F2ren}),
\begin{eqnarray}
\lefteqn{\Bigl( {\rm Term\ 4}\Bigr) \longrightarrow \frac{i 5 \kappa^2 H^6
(a a')^2}{72 \pi^4} \!\times\! \frac14 \mathcal{L}(y) } \nonumber \\
& & \hspace{.8cm} = \frac{i 5 \kappa^2 H^2 (H^2 a a')^2}{288 \pi^4} \Biggl[
{\rm Li}_{2}\Bigl(\frac{y}4\Bigr) + \ln\Bigl(1 \!-\! \frac{y}4\Bigr)
\ln\Bigl(\frac{y}{4}\Bigr) - \frac12 \ln^2\Bigl( \frac{y}{4}\Bigr)^2\Biggr] \; .
\qquad \label{term4}
\end{eqnarray}
The dilogarithm seems intimidating but one sees from expression (\ref{dilog})
that it is analytic at $y=0$, so the $++$ and $+-$ contributions cancel,
\begin{equation}
{\rm Li}_{2}\Bigl(\frac{y_{\scriptscriptstyle ++}}{4}\Bigr) -
{\rm Li}_{2}\Bigl(\frac{y_{\scriptscriptstyle +-}}{4}\Bigr) = 0 \; .
\end{equation}
The logarithm of $1 - \frac{y}{4}$ is also analytic at $y=0$. The only nonzero
contribution comes from the logarithms of $y$,
\begin{eqnarray}
\ln\Bigl( \frac{y_{\scriptscriptstyle ++}}{4}\Bigr) -
\ln\Bigl( \frac{y_{\scriptscriptstyle +-}}{4}\Bigr) & \longrightarrow &
2\pi i \theta(\Delta \eta \!-\! r) \; , \\
\ln^2\Bigl( \frac{y_{\scriptscriptstyle ++}}{4}\Bigr) -
\ln^2\Bigl( \frac{y_{\scriptscriptstyle +-}}{4}\Bigr) & \longrightarrow &
4\pi i \theta(\Delta \eta \!-\! r) \ln\Bigl( -\frac{y}4\Bigr) \; .
\end{eqnarray}
Assembling everything gives,
\begin{eqnarray}
\Bigl( {\rm Term\ 4}\Bigr) & = & -\frac{5 \kappa^2 H^2 (H^2 a a')^2}{144 \pi^3}
\, \theta(\Delta \eta \!-\! r) \Bigl[ \ln\Bigl(1 \!-\! \frac{y}{4}\Bigr) -
\ln\Bigl( -\frac{y}4\Bigr)\Bigr] \; , \\
& = & \frac{\kappa^2 H^2}{72 \pi^3} \times\! -\frac52 (H^2 a a')^2
\theta(\Delta \eta \!-\! r) \ln\Bigl( \frac{4 \!-\! y}{-y}\Bigr) \; , \\
& = & \frac{\kappa^2 H^2}{72 \pi^3} \times\! -\frac52 (H^2 a a')^2
\theta(\Delta \eta \!-\! r) \ln\Bigl( \frac{(\eta \!+\! \eta')^2 \!-\! r^2}{
\Delta \eta^2 \!-\! r^2}\Bigr) \; .
\end{eqnarray}

\subsection{Terms 1-3} \label{spin2-1-3}

What remains is to substitute the various terms from Table~\ref{SKterms}
into the temporal and spatial source integrals on the right hand side of
equation (\ref{oneloop}),
\begin{eqnarray}
\mathcal{S}_{k}(\eta,k) & \equiv & -\partial_0 \! \int \!\! d^4x' \,
iF^{(1)}_{2 , k}(x;x') \partial_0' u_0(\eta',k) e^{-i \vec{k} \cdot
\Delta \vec{x}} \; , \label{Tsource} \\
\overline{\mathcal{S}}_{k}(\eta,k) & \equiv & - k^2 \! \int \!\! d^4x' \,
\Bigl[i F^{(1)}_{2 , k}(x;x') \!+\! iG^{(1)}_{2 , k}(x;x')\Bigr]
u_0(\eta',k) e^{-i \vec{k} \cdot \Delta \vec{x}} \; . \label{Ssource}
\end{eqnarray}
The $k = 1$ terms are local and simple to evaluate,
\begin{eqnarray}
\mathcal{S}_{1}(\eta,k) & = & \frac{85 \kappa^2 H^2}{72 \pi^2} \,
\partial_0 \Bigl[ \ln(a) \partial_0 u_0(\eta,k) \Bigr] \; , \\
\overline{\mathcal{S}}_{1}(\eta,k) & = & \frac{5 \kappa^2 H^2}{72 \pi^2}
\, \ln(a) k^2 u_0(\eta,k) \; .
\end{eqnarray}
Adding the two terms gives,
\begin{equation}
\mathcal{S}_1(\eta,k) + \overline{\mathcal{S}}_1(\eta,k) =
\frac{85 \kappa^2 H^2}{72 \pi^2} \times -i k H a u_0(\eta,k)
+ O\Bigl(\ln(a) \Bigr) \; . \label{source1}
\end{equation}

All the $k > 1$ terms involve the angular integral,
\begin{equation}
\int \!\! d^3x' \, f(r) \, e^{-i \vec{k} \cdot \Delta \vec{x}} = 4\pi
\! \int_{0}^{\infty} \!\!\!\! dr r^2 \, f(r) \frac{\sin(k r)}{k r} \; .
\end{equation}
The $k = 2$ terms contain a radial delta function which immediately
reduces them to single temporal integrations,
\begin{eqnarray}
\mathcal{S}_2(\eta,k) & = & \frac{i \kappa^2 H^2 k^2}{4 \pi^2} \,
\partial_0 \! \int_{\eta_i}^{\eta} \!\!\!\! d\eta' u_0(\eta',k)
\sin(k \Delta \eta) \Bigl[ \ln\Bigl(\frac14 a a'\Bigr) \!+\! \frac13
\!-\! 2 \gamma\Bigr] \; , \qquad \\
\overline{\mathcal{S}}_2(\eta,k) & = & -\frac{\kappa^2 H^2 k^3}{12 \pi^2}
\! \int_{\eta_i}^{\eta} \!\!\!\! d\eta' u_0(\eta',k) \sin(k \Delta \eta)
\Bigl[ \ln\Bigl(\frac14 a a'\Bigr) \!+\! \frac13 \!-\! 2 \gamma\Bigr] \; ,
\end{eqnarray}
where $\eta_i \equiv -H^{-1}$ is the initial time. The core expression
can be reduced to exponential integrals,
\begin{eqnarray}
\lefteqn{ \int_{\eta_i}^{\eta} \!\!\!\! d\eta' e^{-ik \eta'} \!
\sin(k \Delta \eta) \Bigl[ \ln\Bigl(\frac{a a'}4 \Bigr) \!+\! C\Bigr] =
\frac{e^{-ik \eta}}{4 k} \Bigl[1 \!+\! 2 i k
\Delta \eta_i \!-\! e^{2 i k \Delta \eta_i}\Bigr] \Bigl[ \ln\Bigl(
\frac{a}{4}\Bigr) \!+\! C\Bigr] } \nonumber \\
& & \hspace{.8cm} -\frac{e^{-ik \eta}}{2 i H} \Bigl[1 \!-\!
\frac{\ln(a)}{a} \!-\! \frac1{a}\Bigr] + \frac{ \sin(k \eta)}{2 i k} \,
\ln(a) + \frac{e^{i k \eta}}{4 k} \int_{-2k \eta_i}^{-2 k \eta} \!\!
\frac{dt}{t} \Bigl[ e^{i t} \!-\! 1\Bigr] \; . \qquad
\end{eqnarray}
Only the logarithm term on the first line makes a leading order
contribution, and this only for $\mathcal{S}_2(\eta,k)$,
\begin{eqnarray}
\mathcal{S}_2(\eta,k) & = & \frac{\kappa^2 H^2}{16 \pi^2} \times -i k
H a u_0(\eta,k) \Bigl[ e^{2 i k \Delta \eta_i} \!-\! 1 \!-\! 2 i k
\Delta \eta_i\Bigr] + O\Bigl( \ln(a)\Bigr) \; , \label{source2} \qquad \\
\overline{\mathcal{S}}_2(\eta,k) & = & O\Bigl( \ln(a)\Bigr) \; .
\end{eqnarray}

We can also obtain exact results for term 3,
\begin{eqnarray}
\mathcal{S}_3(\eta,k) & = & -\frac{i 5 \kappa^2 H^2}{72 \pi^2}
\partial_0 (\partial_0^2 \!+\! k^2)^2 \! \int_{\eta_i}^{\eta} \!\!\!\!
d\eta' u_0(\eta',k) \nonumber \\
& & \hspace{1cm} \times \int_0^{\Delta \eta} \!\!\!\! dr \, r \sin(kr)
\Biggl\{ \ln\Bigl[ H^2 (\Delta \eta^2 \!-\! r^2)\Bigr] \!-\! 1
\Biggr\} \; , \qquad \\
\overline{\mathcal{S}}_3(\eta,k) & = & \frac{5 \kappa^2 H^2}{72 \pi^2}
k (\partial_0^2 \!+\! k^2)^2 \! \int_{\eta_i}^{\eta} \!\!\!\!
d\eta' u_0(\eta',k) \nonumber \\
& & \hspace{1cm} \times \int_0^{\Delta \eta} \!\!\!\! dr \, r \sin(kr)
\Biggl\{ \ln\Bigl[ H^2 (\Delta \eta^2 \!-\! r^2)\Bigr] \!-\! 1
\Biggr\} \; . \qquad
\end{eqnarray}
Three derivatives can be moved inside the integral because the integrand
vanishes like $\Delta \eta^3 \ln(\Delta \eta)$ for small $\Delta \eta$,
\begin{eqnarray}
\lefteqn{ (\partial_0^2 \!+\! k^2)^2 \! \int_{\eta_i}^{\eta} \!\!\!\!
d\eta' u_0(\eta',k) \! \int_0^{\Delta \eta} \!\!\!\! dr \, r \sin(kr)
\Biggl\{ \ln\Bigl[ H^2 (\Delta \eta^2 \!-\! r^2)\Bigr] \!-\! 1
\Biggr\} } \nonumber \\
& & = 2 k (\partial_0 \!+\! i k) \! \int_{\eta_i}^{\eta} \!\!\!\!
d\eta' u_0(\eta',k) \!\times\! e^{-ik \Delta \eta} \Biggl\{ \int_0^{2 k
\Delta \eta} \!\!\!\!\! dt \Bigl[ \frac{e^{it} \!-\! 1}{t}\Bigr] \!+\! 2
\ln(H \Delta \eta) \Biggr\} \; , \qquad \\
& & = 2 k u_0(\eta,k) \Biggl\{ \int_0^{2 k \Delta \eta_i} \!\!\!\!\! dt
\Bigl[ \frac{e^{it} \!-\! 1}{t}\Bigr] \!+\! 2 \ln(H \Delta \eta_i)
\Biggr\} \; . \qquad \label{reduction}
\end{eqnarray}
Expression (\ref{reduction}) is of order one at late times so neither of
the \# 3 terms contributes at leading order,
\begin{equation}
\mathcal{S}_3(\eta,k) + \overline{\mathcal{S}}_3(\eta,k) = O\Bigl( a^0
\Bigr) \; . \label{source3}
\end{equation}

\subsection{Combining Terms 4-7} \label{spin2-4-7}

The factors of $a a'$ in terms 4-7 suggests very strong contributions, but
it turns out that these cancel when the terms are summed. We first work out
the temporal case. Term 4 requires a partial integration on $r$,
\begin{eqnarray}
\lefteqn{ \mathcal{S}_4(\eta,k) = \frac{i 5 \kappa^2 H^6}{36 \pi^2} \,
\partial_0 \!\! \int_{\eta_i}^{\eta} \!\!\!\! d\eta' (a a')^2 u_0(\eta',k)
\!\! \int_0^{\Delta \eta} \!\!\!\!\!\!\! dr \, r \sin(kr) \ln\Biggl[
\frac{(\eta \!+\! \eta')^2 \!-\! r^2}{\Delta \eta^2 \!-\! r^2} \Biggr] ,} \\
& & \hspace{0cm} = \frac{i 5\kappa^2 H^4}{18 \pi^2} \, \partial_0 \!\!
\int_{\eta_i}^{\eta} \!\!\!\!\! d\eta' \, a a' \ln\Bigl( \frac{a a'}{4}\Bigr)
u_0(\eta',k) \sin(k \Delta \eta) \nonumber \\
& & \hspace{1cm} + \frac{i 5 \kappa^2 H^6}{72 \pi^2} \,
k \partial_0 \!\! \int_{\eta_i}^{\eta} \!\!\!\!\! d\eta' (a a')^2
u_0(\eta',k) \! \int_{0}^{\Delta \eta} \!\!\!\!\! dr \cos(k r) \nonumber \\
& & \hspace{-.7cm} \times \Biggl\{ \! \Bigl[ (\eta \!+\! \eta')^2 \!-\!
r^2\Bigr] \! \ln\Bigl[H^2 \Bigl( (\eta \!+\! \eta')^2 \!-\! r^2\Bigr)\! \Bigr]
\!-\! \Bigl[ \Delta \eta^2 \!-\! r^2 \Bigr] \! \ln\Bigl[H^2 \Bigl( \Delta
\eta^2 \!-\! r^2\Bigr) \! \Bigr] \! \Biggr\} . \qquad \label{S4}
\end{eqnarray}
The surface term of (\ref{S4}) is partially cancelled by
$\mathcal{S}_5(\eta,k)$,
\begin{equation}
\mathcal{S}_5(\eta,k) = -\frac{i 5 \kappa^2 H^4}{36 \pi^2} \, \partial_0
\!\! \int_{\eta_i}^{\eta} \!\!\!\!\! d\eta' \, a a' \Bigl[ \ln\Bigl(
\frac{a a'}{4}\Bigr) \!+\! 2\Bigr] u_0(\eta',k) \sin(k \Delta \eta)
\; . \label{S5}
\end{equation}
The remaining surface term comes from partially integrating
$\mathcal{S}_6(\eta,k)$ on $r$,
\begin{eqnarray}
\lefteqn{ \mathcal{S}_6(\eta,k) = -\frac{i 5 \kappa^2 H^6}{18 \pi^2} \,
\partial_0 \!\! \int_{\eta_i}^{\eta} \!\!\!\! d\eta' (a a')^2 u_0(\eta',k)
\!\! \int_0^{\Delta \eta} \!\!\!\!\!\!\! dr \, \frac{r \sin(k r)}{
(\eta \!+\! \eta')^2 \!-\! r^2} \; , } \\
& & \hspace{-.5cm} = -\frac{i 5\kappa^2 H^4}{36 \pi^2} \, \partial_0 \!\!
\int_{\eta_i}^{\eta} \!\!\!\!\! d\eta' \, a a' \ln\Bigl( \frac{a a'}{4}\Bigr)
u_0(\eta',k) \sin(k \Delta \eta) \nonumber \\
& & \hspace{.3cm} - \frac{i 5 \kappa^2 H^4}{36 \pi^2} \, k \partial_0 \!\!
\int_{\eta_i}^{\eta} \!\!\!\!\! d\eta' a a' u_0(\eta',k) \!
\int_{0}^{\Delta \eta} \!\!\!\!\! dr \cos(k r) \ln\Bigl[ H^2 \Bigl(
(\eta \!+\! \eta')^2 \!-\! r^2\Bigr)\Bigr] \; . \qquad \label{S6}
\end{eqnarray}
Term 7 can be re-expressed by moving a factor of $(\partial_0^2 + k^2)$
inside the integral and then performing some partial integrations on $r$,
\begin{eqnarray}
\lefteqn{ \mathcal{S}_7(\eta,k) = -\frac{i 5 \kappa^2 H^2}{72 \pi^2} \,
\partial_0 \Biggl\{ a (\partial_0^2 \!+\! k^2) \!\! \int_{\eta_i}^{\eta}
\!\!\!\! d\eta' a' u_0(\eta',k) } \nonumber \\
& & \hspace{4.5cm} \times \! \int_0^{\Delta \eta} \!\!\!\!\!\!\!
dr \, r \sin(k r) \Bigl[ \ln\Bigl[ H^2 (\Delta \eta^2 \!-\! r^2)\Bigr]
\!-\! 1 \Bigr] \Biggr\} \; , \qquad \\
& & \hspace{.7cm}=  - \frac{i 5 \kappa^2 H^4}{36 \pi^2} \, k \partial_0
\!\! \int_{\eta_i}^{\eta} \!\!\!\!\! d\eta' a a' u_0(\eta',k) \!
\int_{0}^{\Delta \eta} \!\!\!\!\! dr \cos(k r) \ln\Bigl[ H^2
(\Delta \eta^2 \!-\! r^2)\Bigr] \; . \qquad \label{S7}
\end{eqnarray}
Making some small rearrangements on the sum of (\ref{S4}), (\ref{S5}),
(\ref{S6}) and (\ref{S7}) gives,
\begin{eqnarray}
\lefteqn{ \mathcal{S}_{4-7}(\eta,k) = \frac{i 5 \kappa^2 H^4}{18 \pi^2} \,
k \partial_0 \!\! \int_{\eta_i}^{\eta} \!\!\!\! d\eta' a a' u_0(\eta',k) \!
\int_{0}^{\Delta \eta} \!\!\!\!\! dr \cos(k r)} \nonumber \\
& & \hspace{1cm} \times \Biggl\{-1 + \Bigl( \frac{\Delta \eta^2 \!-\!
r^2}{4 \eta \eta'}\Bigr) \ln\Biggl[ \frac{ (\eta \!+\! \eta')^2 \!-\! r^2}{
\Delta \eta^2 \!-\! r^2}\Biggr] + \frac12 \ln\Biggl[ \frac{ (\eta \!+\!
\eta')^2 \!-\! r^2}{\Delta \eta^2 \!-\! r^2}\Biggr] \Biggr\} . \qquad
\label{S4-7}
\end{eqnarray}
The spatial terms follow similar reductions to give,
\begin{eqnarray}
\lefteqn{ \overline{\mathcal{S}}_{4-7}(\eta,k) =
\frac{5 \kappa^2 H^4}{18 \pi^2} \, k^2 \!\! \int_{\eta_i}^{\eta} \!\!\!\!
d\eta' a a' u_0(\eta',k) \! \int_{0}^{\Delta \eta} \!\!\!\!\! dr \cos(k r)
\Biggl\{1 - 2 \Bigl( \frac{\Delta \eta^2 \!-\! r^2}{
4 \eta \eta'}\Bigr) } \nonumber \\
& & \hspace{1.7cm} + 2 \Bigl( \frac{\Delta \eta^2 \!-\! r^2}{4 \eta \eta'}
\Bigr)^2 \ln\Biggl[ \frac{ (\eta \!+\! \eta')^2 \!-\! r^2}{\Delta \eta^2
\!-\! r^2}\Biggr] - \frac12 \ln\Biggl[ \frac{ (\eta \!+\! \eta')^2 \!-\!
r^2}{\Delta \eta^2 \!-\! r^2}\Biggr] \Biggr\} . \qquad \label{Sbar4-7}
\end{eqnarray}

The representations we have achieved in expressions
(\ref{S4-7}-\ref{Sbar4-7}) are effective for taking the late time limit
because the logarithms vanish like powers of $\eta \eta'$,
\begin{equation}
\ln\Biggl[ \frac{ (\eta \!+\! \eta')^2 \!-\! r^2}{\Delta \eta^2 \!-\! r^2}
\Biggr] = \ln\Biggl[1 \!+\! \frac{4 \eta \eta'}{\Delta \eta^2 \!-\! r^2}
\Biggr] = \frac{4 \eta \eta'}{\Delta \eta^2 \!-\! r^2} - \frac12 \Bigl(
\frac{4 \eta \eta'}{\Delta \eta^2 \!-\! r^2}\Bigr)^2 + \dots \label{expand}
\end{equation}
The expansions of the curly bracketed parts of expressions (\ref{S4-7}) and
(\ref{Sbar4-7}) are,
\begin{equation}
\Biggl\{ \qquad \Biggr\}_{4-7} = \frac56 \Bigl( \frac{4 \eta \eta'}{\Delta
\eta^2 \!-\! r^2}\Bigr)^2 + \dots \quad , \quad
\Biggl\{ \qquad \Biggr\}_{\overline{4}-\overline{7}} = \frac16 \Bigl(
\frac{4 \eta \eta'}{\Delta \eta^2 \!-\! r^2}\Bigr) + \dots \label{toogood}
\end{equation}
These expansions seem to show that (\ref{S4-7}-\ref{Sbar4-7}) are finite
in the late time limit of $\eta \rightarrow 0$, however, this is not quite
correct. When the expansion begins to produce inverse powers of
$(\Delta \eta^2 - r^2)$ it breaks down at the upper limit of the radial
integration, so that the integrals actually grow like $\ln(a)$.

We can obtain analytic forms for the leading growth of
(\ref{S4-7}-\ref{Sbar4-7}) by adding and subtracting to the factor of
$\cos(kr)$,
\begin{equation}
\cos(kr) = \cos(k \Delta \eta) + \Bigl[ \cos(k r) \!-\! \cos(k \Delta \eta)
\Bigr] \; . \label{trick}
\end{equation}
When the square bracketed part of (\ref{trick}) multiplies the curly
bracketed parts of expressions (\ref{S4-7}-\ref{Sbar4-7}) they can be
expanded high enough to give a finite limit for $\eta \rightarrow 0$.
And because the first term of (\ref{trick}) does not depend upon $r$
the radial integration involves only the curly bracketed parts of
(\ref{S4-7}-\ref{Sbar4-7}),
\begin{eqnarray}
\int_{0}^{\Delta \eta} \!\!\!\!\! dr \, \Biggl\{ \qquad \Biggr\}_{4-7}
& = & -\frac13 \Delta \eta - \frac{\eta^2}{3 \eta'} \ln\Bigl(\frac{\eta'}{\eta}
\!-\! 1\Bigr) + \frac{{\eta'}^2}{3 \eta} \ln\Bigl(1 \!-\! \frac{\eta}{\eta'}
\Bigr) \; , \qquad \\
\int_{0}^{\Delta \eta} \!\!\!\!\! dr \, \Biggl\{ \qquad \Biggr\}_{\overline{4}-
\overline{7}} & = & \frac15 \Delta \eta \!-\! \frac2{15}
\frac{\Delta \eta^3}{\eta \eta'} + \Bigl[ \frac{\eta^3}{5 {\eta'}^2} \!-\!
\frac{\eta \Delta \eta^2}{3 {\eta'}^2}\Bigr] \ln\Bigl( \frac{\eta'}{\eta}
\!-\! 1\Bigr) \nonumber \\
& & \hspace{3.5cm} + \Bigl[\frac{\eta' \Delta \eta^2}{3 \eta^2} \!-\!
\frac{{\eta'}^3}{5 \eta^2}\Bigr] \ln\Bigl(1 \!-\! \frac{\eta}{\eta'}\Bigr)
\; . \qquad
\end{eqnarray}
We actually need only the leading behaviors for small $\eta$,
\begin{eqnarray}
\int_{0}^{\Delta \eta} \!\!\!\!\! dr \, \Biggl\{ \qquad \Biggr\}_{4-7}
& = & -\frac12 \eta - \frac{\eta^2}{3 \eta'} \ln\Bigl( \frac{\eta'}{\eta}\Bigr)
+ O\Bigl( \frac{\eta^2}{\eta'}\Bigr) \; , \\
\int_{0}^{\Delta \eta} \!\!\!\!\! dr \, \Biggl\{ \qquad \Biggr\}_{\overline{4}-
\overline{7}} & = & -\frac13 \eta \ln\Bigl( \frac{\eta'}{\eta}\Bigr) + \frac23
\eta + O\Biggl( \frac{\eta^2}{\eta'} \ln\Bigl( \frac{\eta'}{\eta}\Bigr)
\Biggr) \; .
\end{eqnarray}
Substituting in expressions (\ref{S4-7}-\ref{Sbar4-7}) and performing the
temporal integrations gives,
\begin{eqnarray}
\mathcal{S}_{4-7}(\eta,k) & = & -\frac{5 \kappa^2 H^2}{36 \pi^2} \!\times\!
-ik H a \, u_0(\eta,k) + O\Bigl( \ln(a)\Bigr) \; , \label{source4-7} \\
\overline{\mathcal{S}}_{4-7}(\eta,k) & = & O\Bigl( \ln^2(a)\Bigr) \; .
\end{eqnarray}

\subsection{Term 8} \label{spin2-8}

Table~\ref{SKterms} reveals that term 8 has only a spatial part,
\begin{eqnarray}
\overline{\mathcal{S}}_{8}(\eta,k) & \!\!=\!\! &
\frac{\kappa^2 H^4}{12 \pi^2} \, k a (\partial_0^2 \!-\! k^2) \!\!
\int_{\eta_i}^{\eta} \!\!\!\! d\eta' a' u_0(\eta',k) \!
\int_{0}^{\Delta \eta} \!\!\!\!\! dr \, r \sin(k r) \; , \\
& \!\!=\!\! & \frac{\kappa^2 H^4}{6 \pi^2} \, k^2 \!\!
\int_{\eta_i}^{\eta} \!\!\!\! d\eta' a a' u_0(\eta',k) \Delta \eta
\cos(k \Delta \eta) \; , \qquad \\
& \!\!=\!\! & \frac{\kappa^2 H^2}{24 \pi^2} \!\times\! -i k H a u_0(\eta,k)
\!\times\! \Bigl[e^{2 i k \Delta \eta_i} \!-\! 1 \!+\! 2 i k \Delta
\eta_i\Bigr] \!+\! O\Bigl( \ln(a)\Bigr) . \qquad \label{source8}
\end{eqnarray}
One of the peculiarities of this family of exact, de Sitter invariant
gauges is that the spatial part makes leading order contributions
such as (\ref{source8}). In the noncovariant, average gauge only the
temporal part contributes at leading order, and that entirely from the
local term analogous to $\mathcal{S}_1(\eta,k)$ \cite{Wang:2014tza}.

\subsection{Total Leading Spin Two Contribution} \label{total2}

We found leading order contributions from (\ref{source1}), (\ref{source2}),
(\ref{source4-7}) and (\ref{source8}). Their sum is,
\begin{eqnarray}
\lefteqn{\mathcal{S}_{1-7}(\eta,k) + \overline{\mathcal{S}}_{1-8}(\eta,k) }
\nonumber \\
& & \hspace{.5cm} = \frac{\kappa^2 H^2}{48 \pi^2} \!\times\! -ik H a
u_0(\eta,k) \!\times\! \Bigl[45 \!-\! 2 i k \Delta \eta_i \!+\! 5
e^{2 i k \Delta \eta_i}\Bigr] + O\Bigl( \ln^2(a)\Bigr) \; . \qquad
\label{spin2source}
\end{eqnarray}
Substituting (\ref{spin2source}) in the effective mode equation
(\ref{oneloop}) gives the spin two contribution to the one loop mode
function,
\begin{equation}
u_{(1),\rm spin\ 2} = \frac{\kappa^2 H^2}{48 \pi^2}
\frac{i k \ln(a)}{H a} \, u_{(0)}(\eta,k) \!\times\! \Bigl[45 \!-\! 2 i k
\Delta \eta_i \!+\! 5 e^{2 i k \Delta \eta_i}\Bigr] \!+\! O\Bigl(
\frac{\ln^2(a)}{a^2}\Bigr) \; . \quad
\label{spin2mode}
\end{equation}
That compares with the leading result in the noncovariant gauge
\cite{Wang:2014tza},
\begin{equation}
u_{(1),\rm noncovariant} \longrightarrow \frac{\kappa^2 H^2}{48 \pi^2}
 \frac{i k \ln(a)}{H a} \, u_{(0)}(\eta,k) \!\times\! 6 \; .
\label{oldresult}
\end{equation}
Both the new (\ref{spin2mode}) and the old (\ref{oldresult}) results
have the same leading time dependence of $\ln(a)/a$. The signs are also
the same. However, the new result (\ref{spin2mode}) has a different
numerical factor which depends upon the ratio $k/H$.


\section{Spin Zero Contribution}
\label{spin0}

 Here we study the leading order late time one-loop correction of the mode equation
arising from the spin zero part of the graviton propagator. The relevant equation
to solve is Eq.~(\ref{oneloop}), where for $iF^{(1)}(x;x')$ and $iG^{(1)}(x;x')$
one inserts the (retarded part of the) spin-zero contributions~(\ref{F0++}--\ref{G0++}).
In order to simplify the analysis, Eq.~(\ref{oneloop}) can be conveniently written as,
\begin{eqnarray}
(\partial_0^2 \!+\! k^2) u_{(1)}(\eta,k) &=& u_{(0)}(\eta,k)
	\biggl\{ ik \, \partial_0\!\! \int \! d^4x' \, iF_{0}^{(1)}(x;x')
				e^{ik\Delta\eta \!-\! i\vec{k}\cdot\vec{x}}
\label{u1eom}\\
&&	\hskip 1.99cm
	- k^2 \! \int\! d^4x' \, i G_{0}^{(1)}(x;x')
				e^{ik\Delta\eta \!-\! i\vec{k}\cdot\vec{x}} \biggr\}
\,,
\qquad
\nonumber
\end{eqnarray}
where the (retarded) spin-zero structure functions are,
\begin{eqnarray}
\lefteqn{
i F_0^{(1)}(x;x') = \frac{\beta^2\kappa^2H^2}{24\pi^2} \!\times\! \Biggl\{
	- \frac{\ln(a)}{8a} \frac{\partial^2}{H^2}
	\Bigl[ \frac{\delta^4(x\!-\!x')}{a'} \Bigr]
	+ \frac{1}{8} \frac{\partial_0}{aH} \delta^4(x\!-\!x') }
\nonumber \\
& &	\hspace{1cm}
	+ \frac{(\beta\!-\!5)}{12} \ln(a) \delta^4(x\!-\!x')
\label{F0 ret} \\
& &	\hspace{0.5cm}
	+ \frac{1}{128\pi a} \frac{\partial^6}{H^2} \biggl[
	\theta\bigl( \Delta\eta \!-\! \| \Delta\vec{x} \| \bigr) \frac{1}{a'}
	\Bigl[ 1 - \ln \bigl( - \tfrac{H^2}{4} \Delta x^2 \bigr) \Bigr] \biggr]
\nonumber \\
& &	\hspace{0.0cm}
	+ \frac{(\beta\!-\!5)}{192\pi} \partial^4 \biggl[
	\theta\bigl( \Delta\eta \!-\! \| \Delta\vec{x} \| \bigr)
	\Bigl[ 1 \!-\! \ln \bigl( - \tfrac{H^2}{4} \Delta x^2 \bigr) \Bigr]
	- \frac{iH^4}{\pi^2}(aa')^2 \mathcal{N}_F(y) \biggr]
\Biggr\}  ,
\nonumber
\\
\lefteqn{
i G_0^{(1)}(x;x') = \frac{\beta^2\kappa^2H^2}{48\pi^2} \!\times\! \Biggl\{
	\frac{1}{2} \Bigl[ \ln(a) \!-\! 1 \Bigr] \delta^4(x\!-\!x')
	+ \frac{iH^4}{\pi^2}(aa')^2 \mathcal{N}_G(y)  \biggr]}
\nonumber \\
& & \hspace{1cm}
	-\frac{\partial^4}{32\pi}
	\biggl[ \theta\bigl( \Delta\eta \!-\! \| \Delta\vec{x} \| \bigr)
		 \Bigl[ 1 \!-\! \ln \bigl( - \tfrac{H^2}{4} \Delta x^2 \bigr) \Bigr] \biggr]
	\Biggr\} \, ,
\label{G0 ret}
\end{eqnarray}
where the last terms in two expressions above are,
\begin{eqnarray}
	- \frac{i H^4}{\pi^2} (aa')^2 \mathcal{N}_F(y)
	\!\!&=&\!\! \frac{\partial}{\partial\beta} \Biggl\{\!
	- \frac{q_0 B_0}{2} \!\times\! \frac{H^2}{\pi} aa'\, \partial^2
		\theta\bigl( \Delta\eta \!-\! \| \Delta\vec{x} \| \bigr)
\nonumber \\
& &	\hspace{0.75cm}
	- \frac{q_0 A_0}{2} \!\times\! \frac{H^2}{\pi} aa'\ln(aa') \,\partial^2
		\theta\bigl( \Delta\eta \!-\! \| \Delta\vec{x} \| \bigr)
\nonumber \\
& & \hspace{0.75cm}
	- \frac{q_0 A_0}{2} \!\times\!\! \frac{H^2(aa')}{\pi}  \partial^2
		\Bigl[ \ln\bigl( \!- \tfrac{H^2}{4} \Delta x^2 \bigr)
		\theta\bigl( \Delta\eta \!-\! \| \Delta\vec{x} \| \bigr) \Bigr]
\nonumber \\
& & \hspace{0.75cm}
	+ 2q_1B_1 \!\times\! \frac{H^4}{\pi} (aa')^2 \,
		\theta\bigl( \Delta\eta \!-\! \| \Delta\vec{x} \| \bigr)
\nonumber \\
& &	\hspace{0.75cm}
	+ 2q_1A_1 \!\times\! \frac{H^4}{\pi} (aa')^2 \ln(aa') \,
		\theta\bigl( \Delta\eta \!-\! \| \Delta\vec{x} \| \bigr)
\nonumber \\
& & \hspace{0.75cm}
	+2q_1A_1 \!\times\! \frac{H^4}{\pi} (aa')^2
		\ln\bigl( - \tfrac{H^2}{4}\Delta x^2 \bigr) \,
		\theta \bigl( \Delta\eta \!-\! \| \Delta \vec{x} \| \bigr)
\nonumber \\
& &	\hspace{0.75cm}
	+ 2\!\times\!\frac{H^4}{\pi}(aa')^2 S_F(y)\,
		\theta \bigl( \Delta\eta \!-\! \| \Delta \vec{x} \| \bigr)
	\Biggr\}
\, ,
\end{eqnarray}
and
\begin{eqnarray}
	\frac{i H^4}{\pi^2} (aa')^2 \mathcal{N}_G(y)
	\!\!&=&\!\!\! \frac{\partial}{\partial\beta} \Biggl\{\!
	2q_0(2A_0 \!+\! B_0) \!\times\! \frac{(H^2 a a')^2}{\pi}
		\theta\bigl( \Delta\eta \!-\! \| \Delta\vec{x} \| \bigr)
\nonumber \\
& &	\hspace{0.5cm}
	+ 2q_0A_0 \!\times\! \frac{H^4}{\pi}(aa')^2
		\ln\Bigl(\! - \tfrac{H^2}{4} aa' \Delta x^2 \Bigr) \,
		\theta\bigl( \Delta\eta \!-\! \| \Delta\vec{x} \| \bigr)
\nonumber \\
& &	\hspace{0.4cm}
	- 2q_1(2B_1 \!-\! A_1) \!\times\! \frac{H^6}{\pi} (aa')^3
		\Delta x^2 \, \theta\bigl( \Delta\eta \!-\! \| \Delta\vec{x} \| \bigr)
\nonumber \\
& &	\hspace{0.3cm}
	- 4q_1A_1 \!\times\!\! \frac{H^6}{\pi} \!(aa')^3\! \Delta x^2
		\ln\bigl(\! - \tfrac{H^2}{4} aa' \Delta x^2 \bigr) \,
		\!\theta\bigl( \Delta\eta \!-\! \| \Delta\vec{x} \| \bigr)
\nonumber \\
& &	\hspace{0.2cm}
	- 2 \!\times\! \frac{H^4}{\pi} (aa')^2 S_G(y) \,
	\theta\bigl( \Delta\eta \!-\! \| \Delta\vec{x} \| \bigr)  \Biggr\}
\,.
\label{NGR}
\end{eqnarray}
The retarded functions~(\ref{F0 ret}) and (\ref{G0 ret}) are obtained simply by taking the imaginary part 
of $F_0^{(1)}$ and $G_0^{(1)}$ defined in Eqs.~(\ref{F0++}--\ref{G0++}).

 When expressions~(\ref{F0 ret}) and (\ref{G0 ret}) are inserted into~(\ref{u1eom}) one
can express the two principal integrals in terms of 20 relatively simple integrals and six complicated integrals (over
the generalized hypergeometric functions contained in $S_F$ and $S_G$ defined in~(\ref{S_F}--\ref{S_G})) as follows,
\begin{eqnarray}
\!	\int\! d^4x'  e^{ik\Delta\eta - i \vec{k}\cdot\Delta\vec{x}} \, iF_0^{(1)}(x;x')
	\!\!&=&\!\!\frac{\beta^2\kappa^2H^2}{24\pi^2}
	\Biggl\{\! \frac{1}{192} \Bigl[\! -24(I_1\!-\!I_2) + 16(\beta\!-\!5) I_4
\nonumber\\
& &	\hspace{-1.15cm}
	 + \frac{3}{2} (I_5 \!-\! I_6)
	+ (\beta\!-\!5) \bigl( I_7 \!-\! I_8 \bigr) \Bigr]
\nonumber \\
& &	\hspace{-1.6cm}
	+\, \frac{\partial}{\partial\beta} \biggl[\! - \frac{q_0B_0}{2} I_9
	\!-\! \frac{q_0A_0}{2} \bigl( I_{10}\!+\!I_{11} \!+\! I_{12} \bigr)
	\!+\! 2q_1B_1 I_{13}
\nonumber \\
& &	\hspace{-2.05cm}
	+ 2q_1A_1 \bigl( I_{14} \!+\! I_{15} \!+\! I_{16} \bigr)
	\!-\! \frac{(\beta\!-\!4)(\beta\!-\!6)(\beta^2\!-\!20\beta\!+\!40)}{64\!\times\!5!} I_{13}
\nonumber \\
& &	\hspace{-2.5cm}
	+ \frac{(\beta\!-\!4)(\beta\!-\!6)(\beta^2\!-\!12\beta\!+\!40)}{64\!\times\!5!}
	\, \mathcal{I}_{1,0}
		\Bigl( \bigl\{ \tfrac{7}{2}\!+\!b_N, \tfrac{7}{2}\!-\!b_N \bigr\},
		\bigl\{ 6 \bigr\} \Bigr)
\nonumber \\
& &	\hspace{-2.95cm}
	- \frac{\beta(\beta\!-\!4)(\beta\!-\!6)}{8\!\times\!5!} \,
	\mathcal{I}_{2,0}
		\Bigl( \bigl\{ \tfrac{7}{2}\!+\!b_N, \tfrac{7}{2}\!-\!b_N, 1 \bigr\},
		\bigl\{ 6, 2 \bigr\} \Bigr) \biggr]
\label{intFinI} \\
& &	\hspace{-3.4cm}
	+ \frac{5(\beta\!+\!6)(\beta\!-\!4)(\beta\!-\!6)}{16\!\times\!6!} \,
	\mathcal{I}_{3,1}
		\Bigl( \bigl\{ \tfrac{9}{2}\!+\!b_N, \tfrac{9}{2}\!-\!b_N, 1, 1 \bigr\},
		\bigl\{ 7, 2, 2 \bigr\} \Bigr) \biggr] \Biggr\}
\nonumber
\end{eqnarray}
and
\begin{eqnarray}
\!\int\! d^4x'
	e^{ik\Delta\eta - i \vec{k}\cdot\Delta\vec{x}} \, iG_0^{(1)}(x;x')
	\!\!&=&\!\!\frac{\beta^2\kappa^2H^2}{48\pi^2}
	\Biggl\{\! -\frac{1}{2} \bigl( I_3 \!-\! I_4 \bigr)
	\!-\! \frac{1}{32} \bigl( I_7 \!-\! I_8 \bigr)
\nonumber \\
& &	\hspace{-1.2cm}
	+\, \frac{\partial}{\partial \beta} \biggl[
	2q_0 (2A_0 \!+\! B_0) I_{13}
	 \!+\!  2q_0A_0 \bigl( I_{14}  \!+\!  I_{15}  \!+\!  I_{16} \bigr)
\nonumber \\
& &\hspace{-1.6cm}
	-\, 2q_1(2B_1\!-\!A_1) I_{17}
	- 4q_1A_1 \bigl( I_{18} + I_{19} + I_{20} \bigr)
\nonumber \\
& &\hspace{-2.cm}
	+ \frac{\beta(\beta\!-\!4)(\beta\!-\!6)(\beta\!-\!20)}{32 \!\times\! 5!}\, I_{17}
\label{intGinI}  \\
& &\hspace{-2.4cm}
	-\, \frac{(\beta\!-\!4)(\beta\!-\!6)(\beta^2\!-\!12\beta\!+\!40)}{64 \!\times\! 5!}
	\mathcal{I}_{1,1} \,
		\Bigl( \bigl\{ \tfrac{7}{2}\!+\!b_N, \tfrac{7}{2}\!-\!b_N \bigr\} ,
		\bigl\{ 6 \bigr\} \Bigr)
\nonumber \\
& &	\hspace{-2.8cm}
	+ \frac{\beta(\beta\!-\!4)(\beta\!-\!6)}{8 \!\times\! 5!} \,
	\mathcal{I}_{3,1}
	\Bigl( \bigl\{ \tfrac{7}{2}\!+\!b_N, \tfrac{7}{2}\!-\!b_N,1,1 \bigr\},
	\bigl\{ 6,2,2 \bigr\} \Bigr)\nonumber \\
& &	\hspace{-3.2cm}
	- \frac{(\beta\!-\!4)(\beta\!-\!6)(\beta^2\!-\!20\beta\!-\!40)}{64 \!\times\! 5!} \,
	\mathcal{I}_{2,1}
		\Bigl( \bigl\{ \tfrac{7}{2}\!+\!b_N, \tfrac{7}{2}\!-\!b_N, 1 \bigr\} ,
		\bigl\{ 6, 2 \bigr\} \Bigr)
\nonumber\\
& &	\hspace{-3.6cm}
	- \frac{5(\beta\!+\!6)(\beta\!-\!4)(\beta\!-\!6)}{8\!\times\!6!} \,
	\mathcal{I}_{3,2}
	\Bigl( \bigl\{ \tfrac{9}{2}\!+\!b_N, \tfrac{9}{2}\!-\!b_N,1,1 \bigr\},	
	\bigl\{ 7,2,2 \bigr\} \Bigr) \biggr]  \Biggr\}
\,,
\nonumber
\end{eqnarray}
where the 20 simpler integrals are defined as,
\begin{eqnarray}
I_1 & = & \frac{\ln(a)}{a} \int\! d^4x' \,
		e^{ik\Delta\eta - i\vec{k}\cdot\Delta\vec{x}} \,
		\frac{\partial^2}{H^2}  \biggl[ \frac{\delta^4(x\!-\!x')}{a'} \biggr] \ ,
\label{I1def}
\\
I_2 & = & \frac1a\int\! d^4x' \, e^{ik\Delta\eta - i\vec{k}\cdot\Delta\vec{x}} \,
		\frac{\partial_0}{H}  \delta^4(x\!-\!x') \ ,
\label{I2def}
\\
I_3 & = & \int\! d^4x' \, e^{ik\Delta\eta - i\vec{k}\cdot\Delta\vec{x}} \,
		\delta^4(x\!-\!x') \ ,
\label{I3def}
\\
I_4 & = & \ln(a) \int\! d^4x' \, e^{ik\Delta\eta - i\vec{k}\cdot\Delta\vec{x}} \,
		\delta^4(x\!-\!x') \ ,
\label{I4def}
\\
I_5 & = & \frac{1}{\pi a}
		\int\! d^4x' \, e^{ik\Delta\eta - i\vec{k}\cdot\Delta\vec{x}} \,
		\frac{\partial^6}{H^2} \biggl[
		\theta\bigl( \Delta\eta \!-\! \| \Delta\vec{x} \| \bigr) \frac{1}{a'} \biggr] \ ,
\label{I5def}
\\
I_6 & = & \frac{1}{\pi a}
		\int\! d^4x' \, e^{ik\Delta\eta - i\vec{k}\cdot\Delta\vec{x}} \,
		\frac{\partial^6}{H^2} \biggl[
		\theta\bigl( \Delta\eta \!-\! \| \Delta\vec{x} \| \bigr)
		\frac{1}{a'} \ln\bigl( -\tfrac{H^2}{4}\Delta x^2 \bigr) \biggr] \ ,
\label{I6def}
\\
I_7 & = & \frac{1}{\pi} \int\! d^4x' \,
		e^{ik\Delta\eta - i\vec{k}\cdot\Delta\vec{x}} \,
		\partial^4 \theta\bigl( \Delta\eta \!-\! \| \Delta\vec{x} \| \bigr) \ ,
\label{I7def}
\end{eqnarray}
\begin{eqnarray}
I_8 & = & \frac{1}{\pi} \int\! d^4x' \,
		e^{ik\Delta\eta - i\vec{k}\cdot\Delta\vec{x}} \,
		\partial^4 \biggl[ \theta\bigl( \Delta\eta \!-\! \| \Delta\vec{x} \| \bigr)
		\ln\bigl( -\tfrac{H^2}{4}\Delta x^2 \bigr) \biggr] \ ,
\label{I8def}
\\
I_9 & = & \frac{aH^2}{\pi}  \int\! d^4x' \,
		e^{ik\Delta\eta - i\vec{k}\cdot\Delta\vec{x}} \, \partial^2 \Bigl[
		\theta\bigl( \Delta\eta \!-\! \| \Delta\vec{x} \| \bigr) \, a' \Bigr] \ ,
\label{I9def}
\\
I_{10} & = & \frac{a\ln(a)H^2}{\pi} \int\! d^4x' \,
		e^{ik\Delta\eta - i\vec{k}\cdot\Delta\vec{x}} \, \partial^2 \Bigl[
		\theta\bigl( \Delta\eta \!-\! \| \Delta\vec{x} \| \bigr) \, a' \Bigr] \ ,
\label{I10def}
\\
I_{11} & = & \frac{a H^2}{\pi} \int\! d^4x' \,
		e^{ik\Delta\eta - i\vec{k}\cdot\Delta\vec{x}} \, \partial^2 \Bigl[
		\theta\bigl( \Delta\eta \!-\! \| \Delta\vec{x} \| \bigr) \, a' \ln(a') \Bigr] \ ,
\label{I11def}
\\
I_{12} & = & \frac{a H^2}{\pi} \int\! d^4x' \,
		e^{ik\Delta\eta - i\vec{k}\cdot\Delta\vec{x}} \,
		\partial^2 \biggl[ \theta\bigl( \Delta\eta \!-\! \| \Delta\vec{x} \| \bigr) \, a'
		\ln\bigl( - \tfrac{H^2}{4}\Delta x^2 \bigr) \biggr] \ ,
\label{I12def}
\\
I_{13} & = & \frac{a^2 H^4}{\pi} \int\! d^4x' \,
		e^{ik\Delta\eta - i\vec{k}\cdot\Delta\vec{x}} \,
		\theta\bigl( \Delta\eta \!-\! \| \Delta\vec{x} \| \bigr) \, (a')^2 \ ,
\label{I13def}
\\
I_{14} & = & \frac{a^2\ln(a) H^4}{\pi} \int\! d^4x' \,
		e^{ik\Delta\eta - i\vec{k}\cdot\Delta\vec{x}} \,
		\theta\bigl( \Delta\eta \!-\! \| \Delta\vec{x} \| \bigr) \, (a')^2 \ ,
\label{I14def}
\\
I_{15} & = &  \frac{a^2 H^4}{\pi} \int\! d^4x' \,
		e^{ik\Delta\eta - i\vec{k}\cdot\Delta\vec{x}} \,
		\theta\bigl( \Delta\eta \!-\! \| \Delta\vec{x} \| \bigr) \, (a')^2 \ln(a') \ ,
\label{I15def}\\
%
%
I_{16} & = & \frac{a^2 H^4}{\pi} \int\! d^4x' \,
		e^{ik\Delta\eta - i\vec{k}\cdot\Delta\vec{x}} \,
		\theta\bigl( \Delta\eta \!-\! \| \Delta\vec{x} \| \bigr) \, (a')^2
		\ln\bigl( - \tfrac{H^2}{4}\Delta x^2 \bigr) \ ,
\label{I16def}
\\
I_{17} & = & \frac{a^3 H^6}{\pi} \int\! d^4x' \,
		e^{ik\Delta\eta - i\vec{k}\cdot\Delta\vec{x}} \,
		\theta\bigl( \Delta\eta \!-\! \| \Delta\vec{x} \| \bigr) \, (a')^3
		\Delta x^2 \ ,
\label{I17def}
\\
I_{18} & = & \frac{a^3 \ln(a) H^6}{\pi} \int\! d^4x' \,
		e^{ik\Delta\eta - i\vec{k}\cdot\Delta\vec{x}} \,
		\theta\bigl( \Delta\eta \!-\! \| \Delta\vec{x} \| \bigr) \, (a')^3
		\Delta x^2 \ ,
\label{I18def}
\\
I_{19} & = & \frac{a^3 H^6}{\pi} \int\! d^4x' \,
		e^{ik\Delta\eta - i\vec{k}\cdot\Delta\vec{x}} \,
		\theta\bigl( \Delta\eta \!-\! \| \Delta\vec{x} \| \bigr) \, (a')^3
		\ln(a') \Delta x^2 \ ,
\label{I19def}
\\
I_{20} & = & \frac{a^3 H^6}{\pi}\! \int\! d^4x' \,
		e^{ik\Delta\eta - i\vec{k}\cdot\Delta\vec{x}}
		\theta\bigl( \Delta\eta \!-\! \| \Delta\vec{x} \| \bigr)  (a')^3
		\Delta x^2 \ln\bigl( - \tfrac{H^2}{4}\Delta x^2 \bigr) .
\label{I20def}\quad\;\;
\end{eqnarray}
The more complicated integrals over the hypergeometric functions have a general structure,
\begin{eqnarray}
\lefteqn{ \mathcal{I}_{q,N} \Bigl( \bigl\{ \lambda_1, \dots, \lambda_{q+1} \bigr\} ,
	\bigl\{ \sigma_1, \dots, \sigma_q \bigr\} \Bigr) }
\label{hyperIdef}\\
& &	\hspace{0.5cm}
	= \frac{a^2H^4}{\pi} \int\! d^4x' \,
	e^{ik\Delta\eta - i\vec{k}\cdot\Delta\vec{x}} \,
	\theta\bigl( \Delta\eta \!-\! \| \Delta\vec{x} \| \bigr) (a')^2 y^N
	{}_{q+1}F_{q} \Bigl( \bigl\{ \lambda_i \bigr\} , \bigl\{ \sigma_i \bigr\},
		\tfrac{y}{4} \Bigr) \, ,
\nonumber
\end{eqnarray}
and their detailed evaluation in the late time limit is given in
Appendix~A.

 The simpler integrals~(\ref{I1def}--\ref{I20def}) can be all evaluated exactly, and a procedure how to do that 
is briefly outlined in Appendix~B. The results can be
expressed in terms of elementary functions and the following integrals,
\begin{eqnarray}
\mathcal{G}(x) \!& \equiv &\! \int\limits_{0}^{1} \frac{d\tau}{\tau}
	\bigl[ e^{2ix\tau} \!-\! 1 \bigr]
=\big[{\rm ci}(2x)-\gamma_E \!-\! \ln(2x)\big]\!+\!i\Big[{\rm si}(2x)\!+\!\frac\pi 2\Big]
 ,\qquad
\label{funGdef}
\\
\mathcal{M}(x,z) \!& \equiv &\! \int\limits_{0}^{1} \frac{d\tau}{\tau}
	\bigl[ \mathcal{G}(x\tau \!+\! z) \!-\! \mathcal{G}(z) \bigr] \, ,
\label{funMdef}
\\
\mathcal{V}(x,z) \!& \equiv &\! \int\limits_{0}^{1} \frac{d\tau}{\tau}
	\bigl[ e^{2ix\tau} \!-\! 1 \bigr] \mathcal{G}^*(x\tau \!+\! z) \, ,
\label{funVdef}
\end{eqnarray}
where ${\rm ci}$ and ${\rm si}$ are the usual cosine-integral and sine-integral functions defined as,
\begin{eqnarray}
{\rm ci}(z) &=& -\int_z^\infty dt\frac{\cos(t)}{t} = \int_0^zdt\frac{\cos(t)-1}{t}+\gamma_E+\ln(z)
\label{ci function}\\
{\rm si}(z) &=& -\int_z^\infty dt\frac{\sin(t)}{t} = \int_0^z dt\frac{\sin(t)}{t}-\frac\pi 2
\,.
\label{si function}
\end{eqnarray}
 Here it suffices to give the asymptotic form of the simpler integrals~(\ref{I1def}--\ref{I20def}),
which to the relevant order are,
\begin{eqnarray}
I_1 & = &	\mathcal{O}\bigl( \tfrac{\ln(a)}{a} \bigr) \, ,
\label{I1_asymptotic}
\\
I_2 & = &	\mathcal{O}(1) \, ,
\label{I2_asymptotic}
\\
I_3 & = &	\mathcal{O}(1) \, ,
\label{I3_asymptotic}
\\
I_4 & = &	\ln(a) + \mathcal{O}(1) \, ,
\label{I4_asymptotic}
\\
I_5 & = &	\mathcal{O}\bigl( \tfrac{1}{a} \bigr) \, ,
\label{I5_asymptotic}
\\
I_6 & = &	\mathcal{O}(1) \, ,
\label{I6_asymptotic}
\\
I_7 & = &	\mathcal{O}(1) \, ,
\label{I7_asymptotic}
\\
I_8 & = &	\mathcal{O}(1) \,
\label{I8_asymptotic}
\\
I_9 & = &	
	\frac{4iH}{k} \mathcal{G}\bigl( \tfrac{k}{H} \bigr) \!\times\! a + 8\ln(a)
	+ \mathcal{O}(1) \, ,
\label{I9_asymptotic}
\\
I_{10} & = &
	\frac{4iH}{k} \mathcal{G}\bigl( \tfrac{k}{H} \bigr)
		a\ln(a)  \!+\! 8\ln^2(a)
	 \!+\! 8\Bigl[ 1\!+\! \mathcal{G}\bigl( \tfrac{k}{H} \bigr) \Bigr] \ln(a)
	 \!+\! \mathcal{O}\bigl( \tfrac{\ln^2(a)}{a} \bigr) \, ,\qquad
\label{I10_asymptotic}
\end{eqnarray}
\begin{eqnarray}
I_{11} & = &
	\frac{4iH}{k} \mathcal{M}\bigl(\tfrac{k}{H},0\bigr)a
	 \!+\! 4\ln^2(a)  \!+\! 8\ln(a) \!+\! \mathcal{O}(1) \, ,\qquad
\label{I11_asymptotic}
\\
I_{12} & = &
	\frac{4iH}{k} \Bigl[ \mathcal{V}\bigl( \tfrac{k}{H}, 0 \bigr)
		 \!+\! \mathcal{M}\bigl( - \tfrac{k}{H}, 0 \bigr)
		\!-\! 3 \mathcal{M}\bigl( \tfrac{k}{H}, 0 \bigr)
                   \!+\! \mathcal{G}\bigl( \tfrac{k}{H} \bigr) \Bigr] a
\nonumber \\
& &	\hspace{1.cm}		
	- 8\ln^2(a) \!-\! 8\ln(a) \!+\! \mathcal{O}(1) \, ,\quad
\label{I12_asymptotic}
\\
I_{13} & = &
	\frac{2H^2}{k^2} \biggl[ \mathcal{G}\bigl( \tfrac{k}{H} \bigr)\!+\! 2
		\!+\! \frac{iH}{k} \bigl( e^{\frac{2ik}{H}} \!-\! 1 \bigr) \biggr] a^2
	\!+\!\frac{2H^2}{k^2} \biggl[ e^{\frac{2ik}{H}} \!-\!
		1 \!-\! \frac{2ik}{H} \biggr] a
\nonumber \\
& &	\hspace{1.cm}
	+ 4\ln(a) \!+\!\mathcal{O}(1) \, , \qquad
\label{I13_asymptotic}
\\
I_{14} & = &
	\frac{2H^2}{k^2} \biggl[ \mathcal{G}\bigl( \tfrac{k}{H} \bigr) \!+\! 2
		\!+\! \frac{iH}{k} \bigl( e^{\frac{2ik}{H}} \!-\! 1 \bigr) \biggr] a^2 \ln(a)
	\!+\! \frac{2H^2}{k^2} \biggl[ e^{\frac{2ik}{H}} \!-\! 1 \!-\! \frac{2ik}{H} \biggr]a \ln(a)
\nonumber \\
& &	\hspace{1.cm}
	+ 4\ln^2(a) \!+\!2\bigl[ 1 \!+\! 2\mathcal{G}\bigl( \tfrac{k}{H} \bigr) \bigr] \ln(a)
	\!+\! \mathcal{O}\bigl( \tfrac{\ln^2(a)}{a} \bigr) \, ,
\label{I14_asymptotic}
\\
I_{15} & = &
	\frac{2H^2}{k^2} \biggl[ \mathcal{M}\bigl(\tfrac{k}{H}, 0\bigr)
		\!-\! 2\mathcal{G}\bigl( \tfrac{k}{H} \bigr) - 2
		\!+\!  \frac{iH}{k} \bigl( 1 \!-\! e^{\frac{2ik}{H}} \bigr) \biggr] a^2
\label{I15_asymptotic}  \\
& &	\hspace{1.01cm}
	+ \frac{4iH}{k} \biggl[ \mathcal{G}\bigl( \tfrac{k}{H} \bigr) \!+\!  1
	 	\!-\! \frac{iH}{2k} \bigl( 1 \!-\! e^{\frac{2ik}{H}} \bigr) \biggr] a
	\!+\!  2 \ln^2(a) \!+\!  2 \ln(a)\!+\!  \mathcal{O}(1) \, ,
\nonumber
\\
I_{16} & = &
	\frac{2iH^3}{k^3} \biggl\{ 2 e^{\frac{2ik}{H}} \!-\! 2
	\!+\! e^{\frac{2ik}{H}} \mathcal{G}\bigl( -\tfrac{k}{H} \bigr)
	\!-\! \Bigl( 1 \!+\! \frac{4ik}{H} \Bigr) \mathcal{G}\bigl( \tfrac{k}{H} \bigr)
\nonumber \\
& &	\hspace{1.1cm}
	- \frac{ik}{H} \Bigl[ \mathcal{M}\bigl( -\tfrac{k}{H}, 0 \bigr)
		\!-\! 3 \mathcal{M}\bigl( \tfrac{k}{K}, 0 \bigr)
		\!+\!  \mathcal{V}\bigl( \tfrac{k}{H}, 0 \bigr) \Bigr]  \biggr\} a^2
\nonumber \\
& &	\hspace{1.1cm}
	+ \frac{2H^2}{k^2} \biggl\{ 2 e^{\frac{2ik}{H}} \!-\! 2
		\!+\!  e^{\frac{2ik}{H}} \mathcal{G} \bigl( -\tfrac{k}{H} \bigr)
	\!-\! \Bigl( 1 \!+\!  \frac{4ik}{H} \Bigr) \mathcal{G} \bigl( \tfrac{k}{H} \bigr)
	 \biggr\} a \, ,
\nonumber \\
& &	\hspace{1.1cm}
	- 4 \ln^2(a) \!-\! 8 \ln(a)\!+\!  \mathcal{O}(1) \, ,\quad
\label{I16_asymptotic}
\\
I_{17} & = &
	- \frac{12H^4}{k^4} \biggl[ 1 \!+\! \frac{ik}{H}
		\!+\! \frac{ik}{3H} \mathcal{G}\bigl( \tfrac{k}{H} \bigr)
		\!+\! \frac{iH}{2k} \bigl( e^{\frac{2ik}{H}} \!-\! 1 \bigr) \biggr] a^3
\nonumber \\
& &	\hspace{1.3cm}
	+ \frac{12iH^3}{k^3} \biggl[ 1 \!+\! \frac{ik}{3H}
		\!+\! \frac{iH}{2k} \Bigl( 1 \!-\! \frac{2ik}{3H} \Bigr)
		\bigl( e^{\frac{2ik}{H}} \!-\! 1 \bigr) \biggr] a^2
\nonumber \\
& &	\hspace{1.3cm}
	+ \frac{4H^2}{k^2} \biggl[ 1 \!-\! \frac{ik}{H}
		\!+\! \Bigl( 1 \!+\! \frac{iH}{2k} \Bigr) \bigl( e^{\frac{2ik}{H}} - 1 \bigr)
	\biggr] a \!+\! \mathcal{O}(\ln(a)) \, ,
\label{I17_asymptotic}\\
I_{18} & = &
	- \frac{12H^4}{k^4} \biggl[ 1 \!+\! \frac{ik}{H}
		\!+\! \frac{ik}{3H} \mathcal{G}\bigl( \tfrac{k}{H} \bigr)
		\!+\! \frac{iH}{2k} \bigl( e^{\frac{2ik}{H}} \!-\! 1 \bigr) \biggr] a^3 \ln(a)
\nonumber \\
& &	\hspace{1.34cm}
	+ \frac{12iH^3}{k^3} \biggl[ 1 \!+\! \frac{ik}{3H}
		\!+\! \frac{iH}{2k} \Bigl( 1 \!-\! \frac{2ik}{3H} \Bigr)
		\bigl( e^{\frac{2ik}{H}} \!-\! 1 \bigr) \biggr] a^2 \ln(a)
\nonumber \\
& &	\hspace{1.3cm}
	+ \frac{4H^2}{k^2} \biggl[ 1\!-\! \frac{ik}{H}
		\!+\! \Bigl( 1 \!+\! \frac{iH}{2k} \Bigr) \bigl( e^{\frac{2ik}{H}}\!-\! 1 \bigr)
	\biggr] a \ln(a) \!+\! \mathcal{O}(\ln^2(a)) \, ,\quad
\label{I18_asymptotic}
\end{eqnarray}
\begin{eqnarray}
I_{19} \!\!& = &\!\!\!
	- \frac{3iH^5}{k^5} \biggl[ 1 \!+\!  \frac{4ik}{H} - \frac{6k^2}{H^2}
		\!-\! \frac{4k^2}{H^2} \mathcal{G}\bigl( \tfrac{k}{H} \bigr)
	 	\!+\!  \frac{4k^2}{3H^2} \mathcal{M}\bigl( \tfrac{k}{H}, 0 \bigr)
		\!-\! e^{\frac{2ik}{H}} \Bigl( 1 \!+\!  \frac{2ik}{H} \Bigr) \biggr] a^3
\nonumber \\
& &	\hspace{-0.2cm}
	- \frac{3H^4}{k^4} \biggl[ 1 \!+\! \frac{8ik}{3H} \!-\! \frac{10k^2}{3H^2}
		\!-\! \frac{4k^2}{3H^2} \mathcal{G}\bigl( \tfrac{k}{H} \bigr)
		\!-\! e^{\frac{2ik}{H}} \Bigl( 1 \!+\!  \frac{2ik}{3H} \Bigr) \biggr] a^2
\nonumber \\
& &	\hspace{-0.2cm}
	+\frac{iH^3}{k^3} \biggl[ 1\!+\!  \frac{2k^2}{H^2}
		\!+\!  \frac{4k^2}{H^2} \mathcal{G}\bigl( \tfrac{k}{H} \bigr)
		\!-\! e^{\frac{2ik}{H}} \Bigl( 1\!-\! \frac{2ik}{H} \Bigr) \biggr]a
	\!+\!  \mathcal{O}\bigl( \ln^2(a) \bigr)  ,\quad
\label{I19_asymptotic}
\\
I_{20} \!\!& = &\!\!
	\biggl\{\! - \frac{4iH^3}{k^3} \biggl[ \mathcal{V}\bigl( \tfrac{k}{H}, 0 \bigr)
		+ \mathcal{M}\bigl( -\tfrac{k}{H}, 0 \bigr)
		- 3 \mathcal{M}\bigl( \tfrac{k}{H}, 0 \bigr)
		+ 7 \mathcal{G}\bigl( \tfrac{k}{H} \bigr) \biggr]
\nonumber \\
\!\!& &\!\!	\hspace{0.3cm}
	+ \frac{2iH^5}{k^5} \biggl[ 11 \!+\! \frac{16ik}{H} \!-\!
		\frac{10k^2}{H^2} \!+\! 3\Bigl( 1 \!+\! \frac{2ik}{H} \Bigr)
			\mathcal{G}\bigl( \tfrac{k}{H} \bigr)
\nonumber \\
\!\!& &\!\!\!\!\hspace{0.3cm}
	-\, e^{\frac{2ik}{H}} \Bigl( 11 \!+\! \frac{6ik}{H}
		\!+\! 3\mathcal{G}\bigl(-\tfrac{k}{H} \bigr) \Bigr) \biggr] \biggr\} a^3
\nonumber \\
\!\!& &\!\!\!\!\hspace{-0.3cm}
	+\,\frac{2H^4}{k^4} \biggl\{ 11 \!+\! \frac{10ik}{H} \!-\! \frac{6k^2}{H^2}
	\!+\!  \Bigl[ 3 \!+\! \frac{4ik}{H} \!-\! \frac{4k^2}{H^2} \Bigr]
		\mathcal{G}\bigl( \tfrac{k}{H} \bigr)
\nonumber \\
\!\!& &\!\!\!\!\hspace{0.7cm}
	-\, e^{\frac{2ik}{H}} \biggl[ 11 \!+\! \Bigl( 3 \!-\! \frac{2ik}{H} \Bigr)
		\mathcal{G}\bigl( - \tfrac{k}{H} \bigr) \biggr]  \biggr\} a^2
\nonumber \\
\!\!& &\!\!	\hspace{-0.3cm}
	 +\frac{2iH^3}{k^3} \biggl\{
	\!- 4 \!-\! \Bigl( 1 \!+\! \frac{4k^2}{H^2} \Bigr)
		\mathcal{G}\bigl( \tfrac{k}{H} \bigr)
	\!+\!  e^{\frac{2ik}{H}} \biggl[ 4 \!-\! \frac{4ik}{H}
		\!+\! \Bigl( 1 \!-\! \frac{2ik}{H} \Bigr)
		\mathcal{G}\bigl( -\tfrac{k}{H} \bigr) \biggr] \biggr\} a
\nonumber \\
\!\!&&\!\!	\hspace{-0.3cm}
	+\,\mathcal{O}\bigl( \ln^2(a) \bigr)
 \, .
\label{I20_asymptotic}
\end{eqnarray}

Finally, based on the analysis presented in Appendix~A
we arrive at the late time limit of the integrals over hypergeometric functions~(\ref{hyperIdef}),
\begin{eqnarray}
\lefteqn{
\mathcal{I}_{1,0}
	\Bigl( \bigl\{ \tfrac{7}{2} \!+\! b_N, \tfrac{7}{2} \!-\! b_N \bigr\},
		\bigl\{ 6 \bigr\} \Bigr)
	= -\frac{20}{\beta} \frac{iH}{k} \mathcal{G}\bigl( \tfrac{k}{H} \bigr) a
	- \frac{40}{\beta} \Bigl[ 1 \!+\! \frac{8}{(\beta\!-\!4)} \Bigr] \ln(a)
 \, ,\qquad
}\;
\label{I 1 0 : final asymptotic}\\
\lefteqn{
\mathcal{I}_{2,0}
	\Bigl( \bigl\{ \tfrac{7}{2} \!+\! b_N,
		\tfrac{7}{2} \!-\! b_N, 1 \bigr\}, \bigl\{ 6, 2 \bigr\} \Bigr)
	= - \frac{20}{\beta} \frac{iH}{k}
			\mathcal{G}\bigl( \tfrac{k}{H} \bigr) a\ln(a)
}
\nonumber \\
& &	\hspace{-0.3cm}
	- \frac{20}{\beta} \frac{iH}{k} \biggl[
		\Xi\bigl( \tfrac{7}{2}\!+\!b_N, \tfrac{7}{2}\!-\!b_N, 6, \tfrac{3}{2} \bigr)
			\mathcal{G}\bigl( \tfrac{k}{H} \bigr)
		\!-\!2\mathcal{M}\bigl( \tfrac{k}{H}, 0 \bigr)
		\!+\! \mathcal{M}\bigl(\! -\tfrac{k}{H}, 0 \bigr)
		\!+\! \mathcal{V}\bigl( \tfrac{k}{H}, 0 \bigr) \biggr]\!a
\nonumber \\
& &	\hspace{-.2cm}
	-\frac{20}{\beta} \ln^2(a)
	\!+\! \frac{40}{\beta} \biggl[ \frac{8}{(\beta\!-\!4)}
	\!-\! \Xi\bigl( \tfrac{7}{2}\!+\!b_N, \tfrac{7}{2}\!-\!b_N, 6, \tfrac{3}{2} \bigr)
	\!-\! \mathcal{G}\bigl( \tfrac{k}{H} \bigr) \biggr] \ln(a) \, ,
\label{I 2 0 : final asymptotic}
\end{eqnarray}
\begin{eqnarray}
&&\hspace{-0.7cm}
\mathcal{I}_{3,1} \Bigl( \bigl\{ \tfrac{9}{2} \!+\! b_N,
		\tfrac{9}{2} \!-\! b_N, 1, 1 \bigr\}, \bigl\{ 7, 2, 2 \bigr\} \Bigr)
\nonumber \\
& &	\hspace{0.5cm}
	= - \frac{48}{(\beta\!+\!6)} \frac{iH^3}{k^3}
		\biggl[ e^{\frac{2ik}{H}} \!-\! 1\!-\! \frac{2ik}{H}
		\!-\! \frac{ik}{H}\mathcal{G}\bigl( \tfrac{k}{H} \bigr) \biggr] a^2\ln(a)
\nonumber \\
& &	\hspace{0.99cm}
	- \,\frac{48}{(\beta\!+\!6)} \frac{iH^3}{k^3} \biggl\{
	\Xi\bigl( \tfrac{7}{2}\!+\!b_N, \tfrac{7}{2}\!-\!b_N, 6, \tfrac{3}{2} \bigr)
	\biggl[ e^{\frac{2ik}{H}}\!-\! 1\!-\! \frac{2ik}{H}
		\!-\! \frac{ik}{H} \mathcal{G}\bigl( \tfrac{k}{H} \bigr) \biggr]
\nonumber \\
& &	\hspace{0.99cm}	
+\, e^{\frac{2ik}{H}}\Bigl[ 1 \!+\! \mathcal{G}\bigl( -\tfrac{k}{H} \bigr) \Bigr] \!-\! 1 \!+\! \frac{2ik}{H}
\nonumber \\
& &	\hspace{0.99cm}
	+\, \Bigl[ 1 \!+\! \frac{2ik}{H} \Bigr]
			\mathcal{G}\bigl( \tfrac{k}{H} \bigr)
	\!+\!\frac{ik}{H} \biggl[ 2\mathcal{M}\bigl( \tfrac{k}{H}, 0 \bigr)
		\!-\! \mathcal{M}\bigl( - \tfrac{k}{H}, 0 \bigr)
		\!+\! \mathcal{V}\bigl( \tfrac{k}{H}, 0 \bigr)\biggr]
	 \biggr\} a^2
\nonumber \\
& &	\hspace{0.99cm}
	- \frac{48}{(\beta\!+\!6)} \frac{H^2}{k^2} \biggl[
	e^{\frac{2ik}{H}}\!-\! 1\!-\! \frac{2ik}{H} \biggr] a\ln(a)
\nonumber \\
& &	\hspace{0.99cm}
	- \,\frac{48}{(\beta\!+\!6)} \frac{H^2}{k^2} \biggl\{
		\Xi\bigl( \tfrac{7}{2}\!+\!b_N, \tfrac{7}{2}\!-\!b_N, 6, \tfrac{3}{2} \bigr)
	\biggl[ e^{\frac{2ik}{H}} \!-\! 1\!-\! \frac{2ik}{H} \biggr] \!+\!e^{\frac{2ik}{H}}
	\!-\! 1 \!+\! \frac{2ik}{H}
\nonumber \\
& &	\hspace{0.99cm}
	+\, e^{\frac{2ik}{H}} \mathcal{G}\bigl( -\tfrac{k}{H} \bigr)
	\!-\! \biggl[ 1 \!+\! \frac{2ik}{H}\!-\! \frac{10ik}{\beta H} \biggr]
		\mathcal{G}\bigl( \tfrac{k}{H} \bigr)
	 \biggr\}
	\!-\!\frac{48}{(\beta\!+\!6)} \ln^2(a)
\nonumber \\
& &	\hspace{0.99cm}
	-\, \frac{96}{(\beta\!+\!6)} \biggl\{
	\Xi\bigl( \tfrac{7}{2}\!+\!b_N, \tfrac{7}{2}\!-\!b_N, 6, \tfrac{3}{2} \bigr)
	\!+\! \mathcal{G}\bigl( \tfrac{k}{H} \bigr) \!-\! 1 \!-\! \frac{10}{(\beta\!-\!4)} \biggr\}
\label{I 3 1 : final asymptotic}\\
\lefteqn{
\mathcal{I}_{1,1}
	\Bigl( \bigl\{ \tfrac{7}{2} \!+\! b_N, \tfrac{7}{2} \!-\! b_N \bigr\},
		\bigl\{ 6 \bigr\} \Bigr)
	= \frac{320}{\beta(\beta\!-\!4)} \frac{iH}{k}
		\mathcal{G}\bigl( \tfrac{k}{H} \bigr) a \, ,
}
\label{I 1 1 : final asymptotic}\\
\lefteqn{
\mathcal{I}_{2,1}
	\Bigl( \bigl\{ \tfrac{7}{2} \!+\! b_N, \tfrac{7}{2} \!-\! b_N, 1 \bigr\},
		\bigl\{ 6, 2 \bigr\} \Bigr) =
	- \frac{40iH^3}{\beta k^3} \biggl[ e^{\frac{2ik}{H}} \!-\!1 \!-\! \frac{2ik}{H}
		\!-\! \frac{ik}{H} \mathcal{G}\bigl( \tfrac{k}{H} \bigr) \biggr] a^2
}
\nonumber \\
& &	\hspace{1cm}
	- \frac{40H^2}{\beta k^2} \biggl[ e^{\frac{2ik}{H}}\!-\! 1\!-\! \frac{2ik}{H}
	\!+\! \frac{8ik}{(\beta\!-\!4)H} \mathcal{G}\bigl( \tfrac{k}{H} \bigr) \biggr] a
\nonumber\\
\lefteqn{\hspace{-0.cm}
\mathcal{I}_{3,1}
	\Bigl( \bigl\{ \tfrac{7}{2} \!+\! b_N, \tfrac{7}{2} \!-\! b_N, 1, 1 \bigr\},
		\bigl\{ 6, 2, 2 \bigr\} \Bigr)
}
\label{I 3 1B : final asymptotic}\\
 &&\hspace{0.5cm}=
	- \frac{40iH^3}{\beta k^3} \biggl[ e^{\frac{2ik}{H}} \!-\! 1 \!-\! \frac{2ik}{H}
		\!-\! \frac{ik}{H} \mathcal{G}\bigl( \tfrac{k}{H} \bigr) \biggr] a^2 \ln(a)
\nonumber \\
& &	\hspace{1cm}
	-\, \frac{40i H^3}{\beta k^3} \biggl\{
	\Xi \bigl( \tfrac{7}{2} \!+\! b_N, \tfrac{7}{2} \!-\! b_N, 6, \tfrac{3}{2} \bigr)
		\biggl[ e^{\frac{2ik}{H}} \!-\! 1 \!-\! \frac{2ik}{H}
			\!-\! \frac{ik}{H} \mathcal{G}\bigl( \tfrac{k}{H} \bigr) \biggr]
\nonumber \\
& & \hspace{2.88cm}
	+\, e^{\frac{2ik}{H}} \!-\! 1 \!+\! \frac{2ik}{H}
	\!+\! e^{\frac{2ik}{H}} \mathcal{G}\bigl( - \tfrac{k}{H} \bigr)
	\!-\! \Bigl( 1 \!+\! \frac{2ik}{H} \Bigr) \mathcal{G}\bigl( \tfrac{k}{H} \bigr)
\nonumber\\
& &	\hspace{1cm}
	+ \frac{ik}{H} \biggl[ 2\mathcal{M}\bigl( \tfrac{k}{H}, 0 \bigr)
		\!-\! \mathcal{M}\bigl( - \tfrac{k}{H}, 0 \bigr)
		\!-\! \mathcal{V}\bigl( \tfrac{k}{H}, 0 \bigr) \biggr]
	\biggr\} a^2
\nonumber
\end{eqnarray}
\begin{eqnarray}
\nonumber \\
& &	\hspace{1cm}
	- \frac{40 H^2}{\beta k^2} \biggl[ e^{\frac{2ik}{H}} \!-\! 1 \!-\! \frac{2ik}{H} \biggr] a\ln(a)
\nonumber \\
& & \hspace{1cm}
	- \frac{40 H^2}{\beta k^2} \biggl\{
	\Xi\bigl( \tfrac{7}{2} \!+\! b_N, \tfrac{7}{2} \!-\! b_N, 6, \tfrac{3}{2} \bigr)
		\biggl[ e^{\frac{2ik}{H}} \!-\! 1 \!-\! \frac{2ik}{H} \biggr]
		\!+\! e^{\frac{2ik}{H}}\!-\! 1 \!+\! \frac{2ik}{H}
\nonumber \\
& &	\hspace{2.7cm}
	- \Bigl( 1 \!-\! \frac{2ik\beta}{(\beta\!-\!4)H} \Bigr) \mathcal{G}\bigl( \tfrac{k}{H} \bigr)
	\!+\! e^{\frac{2ik}{H}} \mathcal{G}\bigl( - \tfrac{k}{H} \bigr)
\biggr\} a
\nonumber\\
\lefteqn{
\mathcal{I}_{3,2}
	\Bigl( \bigl\{ \tfrac{9}{2} \!+\! b_N, \tfrac{9}{2} \!-\! b_N, 1, 1 \bigr\},
		\bigl\{ 7, 2, 2 \bigr\} \Bigr)
}
\label{I 3 2 : final asymptotic} \\
 &&\hspace{0.5cm}
    = \frac{48iH^5}{(\beta\!+\!6)k^5} \biggl[
		3 e^{\frac{2ik}{H}} \!-\! 3 \!-\! \frac{6ik}{H} \!+\! \frac{6k^2}{H^2}
		\!+\! \frac{2k^2}{H^2} \mathcal{G}\bigl( \tfrac{k}{H} \bigr) \biggr] a^3 \ln(a)
\nonumber\\
& &	\hspace{0.5cm}
	+\, \frac{24 H^5}{(\beta\!+\!6)k^5} \biggl\{\!
	6i  \Xi\bigl( \tfrac{9}{2}\!+\!b_N, \tfrac{9}{2}\!-\!b_N, 7, \tfrac{3}{2} \bigr)
	\biggl[ e^{\frac{2ik}{H}} \!-\! 1 \!-\! \frac{2ik}{H} \!+\! \frac{2k^2}{H^2}
		\!+\! \frac{2k^3}{3H^3} \mathcal{G}\bigl( \tfrac{k}{H} \bigr) \biggr]
\nonumber \\
& &	\hspace{2.5cm}
	+ 19i \bigl( e^{\frac{2ik}{H}} \!-\! 1 \bigr) \!+\! \frac{20k}{H}
		\!-\! \frac{6k}{H} e^{\frac{2ik}{H}} \!+\! \frac{2ik^2}{H^2}
		\!+\! 6i e^{\frac{2ik}{H}} \mathcal{G}\bigl( - \tfrac{k}{H} \bigr)
\nonumber \\
& &	\hspace{2.5cm}
	+ \Bigl(\! -6i \!+\! \frac{12k}{H} \!+\! \frac{16i k^2}{H^2} \Bigr) \mathcal{G}\bigl( \tfrac{k}{H} \bigr)
\nonumber \\
& &	\hspace{2.5cm}
	+ \frac{4ik^2}{H^2} \Bigl[ \mathcal{M}\bigl( - \tfrac{k}{H}, 0 \bigr)
		 \!-\! 2 \mathcal{M}\bigl( \tfrac{k}{H}, 0 \bigr)
		\!+\! \mathcal{V}\bigl( \tfrac{k}{H}, 0 \bigr) \Bigr]
	\biggr\} a^3
\nonumber \\
& &	\hspace{0.5cm}
	- \frac{48 H^4}{k^4(\beta\!+\!6)} \biggl[
		3 \!+\!\frac{4ik}{H} \!-\!\frac{2k^2}{H^2}
		\!+\! e^{\frac{2ik}{H}} \Bigl( \!-3 \!+\! \frac{2ik}{H} \Bigr) \biggr] a^2\ln(a)
\nonumber \\
& &	\hspace{0.5cm}
	+ \frac{24H^4}{(\beta\!+\!6)k^4} \biggl\{\!
		2\, \Xi\bigl( \tfrac{9}{2}\!+\!b_N, \tfrac{9}{2}\!-\!b_N, 7, \tfrac{3}{2} \bigr)
		\biggl[\!-3  \!-\! \frac{4ik}{H} + \frac{2k^2}{H^2}
		\!+\! e^{\frac{2ik}{H}} \Bigl(\!3 \!-\! \frac{2ik}{H} \Bigr) \biggr]
\nonumber \\
& &	\hspace{2.4cm}
	+ 19\bigl( e^{\frac{2ik}{H}} \!-\! 1 \bigr) \!-\! \frac{12ik}{H} \!+\! \frac{2k^2}{H^2}
	\!-\! \frac{2ik}{H} e^{\frac{2ik}{H}}
	\!+\! 2 e^{\frac{2ik}{H}} \Bigl(\! 3 \!-\! \frac{2ik}{H} \Bigr)
		\mathcal{G}\bigl(\! -\tfrac{k}{H} \bigr)
\nonumber \\
& &	\hspace{2.4cm}
	+ 2 \Bigl(\! -3 \!-\! \frac{4ik}{H} \!+\! \frac{2k^2}{H^2} \Bigr) \mathcal{G}\bigl( \tfrac{k}{H} \bigr)
	\biggr\} a^2
\nonumber \\
& &	\hspace{0.5cm}
	+ \frac{48iH^3}{(\beta\!+\!6)k^3} \biggl[
		1 \!+\! \frac{2k^2}{H^2} \!-\! e^{\frac{2ik}{H}} \Bigl( 1 \!-\! \frac{2ik}{H} \Bigr) \biggr]
		a\ln(a)
\nonumber \\
& &	\hspace{0.5cm}
	- \frac{24H^3}{(\beta\!+\!6)k^3} \biggl\{
		2i\, \Xi\bigl( \tfrac{9}{2}\!+\!b_N,\tfrac{9}{2}\!-\!b_N, 7, \tfrac{3}{2} \bigr)
			\biggl[ e^{\tfrac{2ik}{H}} \Bigl( 1 \!-\! \frac{2ik}{H} \Bigr)
			 \!-\! 1 \!-\! \frac{2k^2}{H^2} \biggr]
\nonumber \\
& &	\hspace{2.5cm}
		+ e^{\frac{2ik}{H}} \Bigl( 7i \!+\! \frac{6k}{H} \Bigr)
	\!-\! 7i \!-\! \frac{2ik^2}{H^2}
	\!+\! 2i\, e^{\frac{2ik}{H}} \Bigl( 1 \!-\! \frac{2ik}{H} \Bigr)
		\mathcal{G}\bigl( -\tfrac{k}{H} \bigr)
\nonumber \\
& &	\hspace{2.5cm}
	- 2i \biggl[ 1\!+\! \frac{2(\beta^2\!-\!4\beta\!-\!40)}{\beta(\beta\!-\!4)} \frac{k^2}{H^2} \biggr]
		\mathcal{G}\bigl( \tfrac{k}{H} \bigr) \biggr\} a \ ,
\nonumber
\end{eqnarray}
\begin{eqnarray}
& &	\hspace{0cm}
\end{eqnarray}
where we have defined
\begin{equation}
\Xi(\alpha_1,\alpha_2,\beta_1,\beta_2)
	= \psi(\alpha_1 \!-\! 1) + \psi(\alpha_2 \!-\! 1)
	- \psi(\beta_1 \!-\! 1) - \psi(\beta_2 \!-\! 1) - 2\ln(2) \, .
\end{equation}

Now, plugging in all the integrals (\ref{I1_asymptotic}--\ref{I20_asymptotic})
and (\ref{I 1 0 : final asymptotic}--\ref{I 3 2 : final asymptotic}) 
into expressions (\ref{intFinI})
and (\ref{intGinI}) gives
\begin{eqnarray}
& & \int \! d^4x'\, e^{ik\Delta\eta - i \vec{k}\cdot \Delta\vec{x}} \, i \, F_R^0(x;x') = 0
	\qquad \text{to order $\ln(a)$} \, ,
\\
& & \int \! d^4x'\, e^{ik\Delta\eta - i \vec{k}\cdot \Delta\vec{x}} \, i \, G_R^0(x;x') = 0
	\qquad \text{to order $a$} \, .
\end{eqnarray}
In view of Eq.~(\ref{u1eom}),
this then implies that there is {\it no} one-loop correction from gravitons
that contributes at the leading order as $\propto [\ln(a)/a]u_{(0)}(\eta,k)$
(see Eq.~(\ref{spin2mode})) to the photon wave function $u_{(1)}(\eta,k)$.

This completes the analysis of the graviton induced one-loop correction to the photon wave function on de Sitter.
This analysis shows that, at late times, the leading contribution comes entirely from the spin-two part of the
graviton propagator (\ref{spin2mode}), implying that our result is independent on the graviton gauge parameter $b$
(or, equivalently, on the parameter $\beta=(4b-2)/(b-2)$) for $b>2$.


\section{Discussion}
\label{discuss}

Inflation creates an ensemble of gravitons. We have studied the
effects that these gravitons have on the propagation of a spatial plane 
wave photon. What we find is that the one loop electric field strength 
grows, relative to the tree order result, by an amount which eventually 
becomes nonperturbatively strong,
\begin{equation}
F_{0i}^{(1)} \longrightarrow \frac{\kappa^2 H^2}{48 \pi^2} \, \ln(a) 
\Bigl[45 \!-\! \frac{2i k}{H} \!+\! 5 e^{2i k/H}\Bigr] \!\times\!
F_{0i}^{(0)} \; .
\label{ourresult}
\end{equation}
This field comes entirely from the effect of the spin-two part 
of the graviton propagator on the photon mode function~(\ref{spin2mode}),
$F_{0i}^{(1)}=\partial_0 u_{(1),\rm spin\ 2}(\eta,k)\epsilon^i(\vec k\,) \times e^{i\vec k\cdot \vec x}$.
Note that~(\ref{spin2mode}) implies that there is no secular growth 
in the magentic field during inflation.
The physical interpretation of the result~(\ref{ourresult}) seems to be that a photon
is scattered more and more as it propagates through the ensemble of
horizon-scale gravitons created by inflation. The photon's physical 
3-momentum redshifts like $\tfrac{k}{a(t)}$, whereas inflationary 
particle production continually replenishes the supply of gravitons 
with physical 3-momentum $\tfrac{k}{a(t)} \sim H$. The spin-spin coupling allows 
these gravitons to interact with the redshifting photon to arbitrarily 
late times. A scattering is rare --- because quantum gravity is weak, 
even at inflationary scales --- but it essentially always adds to the 
photon's 3-momentum, and therefore increases its electric field 
strength.

No one doubts that an ensemble of gravitons on flat space would 
scatter a photon --- indeed, this is the basis of attempts to detect 
gravitational waves by pulsar timing --- so there should be no surprise
that it happens on de Sitter. However, we do need to infer the effect 
in a way which does not depend upon the choice of graviton gauge. 
Checking this was one of the primary motivations for our work, and we
did check that the secular growth factor (\ref{ourresult}) has no
dependence on the parameter $b$ which characterizes a general, exact,
de Sitter invariant gauge \cite{Mora:2012zi}. The vacuum polarization
in this gauge depends massively upon $b$ \cite{Glavan:2015ura}, yet
we saw in section 4 that none of the $b$-dependent terms contribute
to the secular growth (\ref{ourresult}). That supports the secular 
gauge independence conjecture \cite{Miao:2012xc}. 

Unfortunately, our result (\ref{ourresult}) is not the same as was 
previously obtained \cite{Wang:2014tza} in a noncovariant, average 
gauge \cite{Tsamis:1992xa}. It has the same sign and spacetime 
dependence, but the noncovariant average gauge has the factor $45
+ 2ik/H + 5 e^{2ik/H}$ replaced by just $6$. This may mean that the
secular gauge independence conjecture is wrong. However, another 
possibility is that there is an obstacle to imposing the de Sitter 
breaking, average gauge, just as there has already been shown to be
an obstacle to imposing de Sitter invariant, average gauges 
\cite{Miao:2009hb}.

 Since the the electric field~(\ref{ourresult}) exhibits a secular growth,
 it will become large during inflation if inflation lasts long enough.
The question that naturally arises is whether that field can give rise 
to magnetogenesis by postinflationary dynamics. 
The crucial difference between the graviton effect considered here
and the effect induced by (light or massless minimally coupled) charged
scalars is in that charged scalar fluctuations generate 
a photon mass~\cite{Prokopec:2002uw,Prokopec:2002jn,Prokopec:2003iu,Prokopec:2003tm,Prokopec:2007ak},
while graviton fluctuations only modify the wave function.
The postinflationary magnetogenesis crucially depends on 
the photon mass~\cite{Davis:2000zp,Prokopec:2003bx,Prokopec:2004au},
since it is the photon mass that is responsible for generation of modified 
electric and magnetic field spectra. In the case under consideration however, 
the electric field~(\ref{ourresult}) gets amplified,
but the (electric and magnetic)
field spectra remain unmodified during inflation, meaning that (on super-Hubble scales) 
they are suppressed as $\propto k^4$. 
This then implies that postinflationary physics will transfer the energy 
from the electric to the magnetic field, reaching eventually equipartition (if that is not prevented by 
large conductivity that could be generated during postinflationary thermalization). 
The result of that process will be tiny primordial magnetic fields on cosmological scales,
but larger magnetic fields on small scales, of the order of meter and larger 
(recall that the comoving Hubble scale at the end of  inflation at the grand unified scale
corresponds to about 1 meter today).

\bigskip\medskip
\centerline{\bf Acknowledgements}
\medskip

We are grateful for conversation and correspondence on this subject
with S. Deser. This work was partially supported by Polish National
Science Centre grant 2014/14/E/ST9/00152, by Taiwan MOST grant
103-2112-M-006-001-MY3, by the D-ITP consortium, a program of the
NWO that is funded by the Dutch Ministry of Education, Culture and
Science (OCW), by NSF grant PHY-1506513, and by the Institute for
Fundamental Theory at the University of Florida. We also thank the
University of Utrecht for its hospitality during the final work on
this project.


\section*{Appendix A: The integrals over hypergeometric functions  
}
\label{app: Integrals over hypergeometric functions}

Here we sketch important steps in the computation of the integrals~(\ref{hyperIdef}),
\begin{eqnarray}
&&\hskip -1cm
\mathcal{I}_{q,N} \Bigl( \bigl\{ \lambda_i \bigr\}, \bigl\{ \sigma_i \bigr\} \Bigr)
\label{I qN def}\\
\hskip 0.3cm	 &&=\, \frac{a^2H^4}{\pi} \! \int\! d^4x'\,
	e^{ik\Delta\eta - i\vec{k}\cdot\Delta\vec{x}}\,
	\theta\bigl( \Delta\eta \!-\! \| \Delta\vec{x} \| \bigr)
	(a')^2 y^N \, {}_{q+1}F_q
		\Bigl( \bigl\{ \lambda_i \bigr\}, \bigl\{ \sigma_i \bigr\}, \tfrac{y}{4} \Bigr)
	\, ,
\nonumber
\end{eqnarray}
in the late time
limit, $a\!\gg\!1$ and $k/(aH)~\!\ll\!~1$. Here the following quantities are defined,
\begin{equation}
\Delta\eta = \eta \!-\! \eta' \, , \qquad
	\Delta \vec{x} = \vec{x} \!-\! \vec{x}^{\,\prime} \, ,\qquad
	y = H^2 aa' \bigl[ \| \Delta\vec{x} \,\|^2 \!-\! \Delta\eta^2 \,\bigr] \, .
\end{equation}

First we integrate over the spatial angular coordinates,
where we introduced $\vec{r}\!=\!\Delta\vec{x}$ and $r\!=\! \| \vec{r}\, \|$
(so now $y\!=\! H^2 aa' [r^2 \!-\! \Delta\eta^2]$),
\begin{eqnarray}
\mathcal{I}_{q,N} = 4a^2H^4\!\!\! \int\limits_{-1/H}^{-1/(aH)} \!\! d\eta' \,
e^{ik\Delta\eta}\, (a')^2\!
	\int\limits_{0}^{\Delta\eta} \! dr r^2  \frac{\sin(kr)}{kr} \,
	y^N \, {}_{q+1}F_q
	\Bigl( \bigl\{ \lambda_i \bigr\}, \bigl\{ \sigma_i \bigr\}, \tfrac{y}{4} \Bigr) \, .
\nonumber\\
\label{I qN def:2}
\end{eqnarray}
Next we switch to a dimensionless time integration variable $\tau$,
\begin{equation}
\tau \equiv - H\eta' \, ,\qquad \Delta\tau \equiv H\Delta\eta = \tau -\frac{1}{a} \, ,
\end{equation}
and to a dimensionless radial integration variable  $\rho$,
\begin{equation}
r = \rho \Delta\eta = \frac{\rho \Delta\tau}{H} \, ,
\end{equation}
and define the shorthand notation for a dimensionless momentum,
\begin{equation}
K = \frac{k}{H} \, .
\end{equation}
This turns the integral~(\ref{I qN def:2}) into
\begin{eqnarray}
\mathcal{I}_{q,N} \!&=&\! (-1)^N 4a^{2+N} \!
	\int\limits_{1/a}^{1} \! d\tau \,
		e^{iK\Delta\tau}\, \tau^{-2-N} (\Delta\tau)^{3+2N}
\label{I qN def:3}\\
& &	\hspace{0.cm}
	\!\!\times\! \int\limits_{0}^{1} \! d\rho\, \rho^2 \,
	(1\!-\!\rho^2)^N \, \frac{\sin(K\rho\Delta\tau)}{K\rho\Delta\tau} \,
	{}_{q+1}F_q
		\Bigl( \bigl\{ \lambda_i \bigr\}, \bigl\{ \sigma_i \bigr\},
		\!-\tfrac{a(\Delta\tau)^2}{4\tau} [1\!-\!\rho^2] \Bigr) . \quad
\nonumber
\end{eqnarray}
Trigonometric functions are uniformly convergent on the whole real line, so we may
expand them in a power series,
and interchange the summation and integration operations
in~(\ref{I qN def:3}). The power series is
\begin{equation}
\frac{\sin(K\rho\Delta\tau)}{K\rho\Delta\tau}
	= \sum_{n=0}^{\infty} \frac{(-1)^n}{(2n\!+\!1)!} (K\rho\Delta\tau)^{2n} \, ,
\nonumber
\end{equation}
and we can write the integral~(\ref{I qN def:3}) as
\begin{eqnarray}
\lefteqn{
\mathcal{I}_{q,N} = 4(-1)^N a^{2+N}
	\sum_{n=0}^{\infty} \frac{(-1)^n}{(2n\!+\!1)!} K^{2n}\!\!
	\int\limits_{1/a}^{1}\! d\tau\, e^{iK\Delta\tau}
		\tau^{-2-N} (\Delta\tau)^{3+2N+2n}
}
\label{I qN def:4} \\
& &	\hspace{2.6cm}
	\times \int\limits_{0}^{1}\! d\rho\, \rho^{2+2n} (1 \!-\! \rho^2)^N \,
	{}_{q+1}F_q
		\Bigl( \bigl\{ \lambda_i \bigr\}, \bigl\{ \sigma_i \bigr\},
		- \tfrac{a(\Delta\tau)^2}{4\tau} [1\!-\!\rho^2] \Bigr) \, .
\nonumber
\end{eqnarray}
Next, making a substitution of variable,
\begin{equation}
\nu \!=\! 1 \!-\! \rho^2
\,,
\end{equation}
puts the integral~(\ref{I qN def:4}) in the form
\begin{eqnarray}
\lefteqn{
\mathcal{I}_{q,N} = 2(-1)^N a^{2+N}
	\sum_{n=0}^{\infty} \frac{(-1)^n}{(2n\!+\!1)!} K^{2n}\!\!
	\int\limits_{1/a}^{1}\! d\tau\, e^{iK\Delta\tau}
		\tau^{-2-N} (\Delta\tau)^{3+2N+2n}
}
\label{I qN def:5} \\
& &	\hspace{3.2cm}
	\times \int\limits_{0}^{1} d\nu
		(1\!-\!\nu)^{n+\frac{1}{2}} \, \nu^N \,
	{}_{q+1}F_q
		\Bigl( \bigl\{ \lambda_i \bigr\}, \bigl\{ \sigma_i \bigr\},
		- \tfrac{a(\Delta\tau)^2}{4\tau} \nu \Bigr) \, ,
\nonumber
\end{eqnarray}
where now the integral over $\nu$ can be done exactly,~\footnote{
One can make use of the integral 7.512.12 from~\cite{Gradshteyn:2007}. However,
that integral requires $aa' \Delta\tau^2/4\!<\!1$, which is equivalent to $\tfrac{a'}{a}+\tfrac{a}{a'}<6$,
which is broken at early times when $t^\prime$ is much before $t$ (more precisely when $a'/a<3-\sqrt{8}$).
Since the result of integration is proportional to a hypergeometric function, it is reasonable to assume that
the result applies in the whole region of integration 
in the sense that the hypergeometric function in~(\ref{I qN def:5}) is defined on the whole complex plane (except
on the cuts).
}
\begin{eqnarray}
\mathcal{I}_{q,N} \!\!&=&\!\! 2 (-1)^N (N!)
	\sum_{n=0}^{\infty} \frac{(-1)^n}{(2n\!+\!1)!} K^{2n}
	\frac{ \Gamma \bigl( n \!+\! \frac{3}{2} \bigr)}
		{\Gamma\bigl( n \!+\! N \!+\! \frac{5}{2} \bigr)}
	a^{2+N}\!\!\int\limits_{1/a}^{1} \! d\tau\,
	e^{iK\Delta\tau} \tau^{-2-N} \quad
\nonumber \\
\!\!&\times &\!\!\!
	 (\Delta\tau)^{3+2N+2n}\, {}_{q+2}F_{q+1}
		\Bigl( \bigl\{ \lambda_i, 1\!+\!N \bigr\},
			\bigl\{ \sigma_i, n \!+\! N \!+\! \tfrac{5}{2} \bigr\},
			- \tfrac{a(\Delta\tau)^2}{4\tau} \Bigr)  . \qquad
\label{I intermediate}
\end{eqnarray}

Next we start approximating the integral under the sum in~(\ref{I intermediate}), which we denote as,
\begin{eqnarray}
\lefteqn{
\mathcal{J}_{q,N}^n = a^{2+N} \!
	\int\limits_{1/a}^{1} \! d\tau\,
	e^{iK\Delta\tau} \tau^{-2-N} (\Delta\tau)^{3+2N+2n}
}
\label{I intermediate:n} \\
& &	\hspace{2.1cm}
	\times \, {}_{q+2}F_{q+1}
		\Bigl( \bigl\{ \lambda_i, 1\!+\!N \bigr\},
			\bigl\{ \sigma_i, n \!+\! N \!+\! \tfrac{5}{2} \bigr\},
			- \tfrac{a(\Delta\tau)^2}{4\tau} \Bigr) \, .
\nonumber
\end{eqnarray}
We do not know how to evaluate the full integral, instead we seek to find
the late time behavior for $a\!\gg\!1$ and $k/(aH)\ll 1$. In particular, we want to isolate
the late time growing terms up to order $\ln(a)$
or $a$, depending whether it appears in the integral over $F$ or over $G$, respectively.

Let us now make a variable substitution,
\begin{equation}
\tau = \frac{t}{a} \;\Rightarrow\; t = \frac{a}{a'}\, ,
\end{equation}
which puts the integral~(\ref{I intermediate:n}) into the form
\begin{eqnarray}
\lefteqn{
\mathcal{J}_{q,N}^{n} = a^{-2n} \!
	\int\limits_{1}^{a} \! dt\,
	e^{\frac{iK(t-1)}{a}} t^{-2-N} (t\!-\!1)^{3+2N+2n}
}
\nonumber \\
& &	\hspace{2.cm}
	\times \, {}_{q+2}F_{q+1}
		\Bigl( \bigl\{ \lambda_i, 1\!+\!N \bigr\},
			\bigl\{ \sigma_i, n \!+\! N \!+\! \tfrac{5}{2} \bigr\},
			- \tfrac{(t-1)^2}{4t} \Bigr) \, . \qquad
\label{J intermediate}
\end{eqnarray}
Because of the factor $a^{-2n}$ outside, we need to identify only the
the contributions to the remaining integral that grow as $a^{2n}$ or faster
in the late time limit. We will do that by approximating the integrand by a
much simpler function, which we will be able to integrate over. It is of no
relevance if we retain some terms that contribute to subleading orders at
late times ({\it i.e.} that grow slower than $a^{2n}$),
since in the end we will neglect them anyway. What is important is
just for the new approximated integrand to capture correctly the relevant
late time terms.

The smallest parameter in the first set of parameters of the hypergeometric function
in the integrand can be 1. Therefore, the leading behavior of the hypergeometric
function for large arguments is $\sim t^{-1}$ (and possibly times some integer
powers of $\ln(t)$, which does not change the argument). Therefore, for large
$t$, the leading behavior of the integrand is $\sim t^{N+2n}$
(the phase factor $\exp[iK(t-1)/a]$
does not change the argument either). This means that the
leading late time behavior is (not counting the powers of logarithms),
\begin{equation}
\mathcal{J}_{q,N}^n \sim a^{-2n} \int^a dt\, t^{N+2n} \sim a^{1+N} \, ,
\end{equation}
and is independent of parameter $n$. Therefore, by extracting the first $N\!+\!2$ (recall that $N=0,1,2$)
terms in the asymptotic expansion of the hypergeometric function we obtain
all the relevant contributions to $\mathcal{J}_{q,N}^n$ at late times. The asymptotic
expansion of hypergeometric functions will take the form
(16.11.6 and 16.11.2 from \cite{DLMF}),
\begin{eqnarray}
\lefteqn{
{}_{q+2}F_{q+1}
		\Bigl( \bigl\{ \lambda_i, 1\!+\!N \bigr\},
			\bigl\{ \sigma_i, n \!+\! N \!+\! \tfrac{5}{2} \bigr\},
			- \tfrac{(t-1)^2}{4t} \Bigr)
}
\nonumber \\
& &	\hspace{2.cm}
	\approx \sum_{l=1}^{N+1+s_*} t^{-l} \times
		c_{q,N}^{n,l}(\lambda_i, \sigma_i, \ln(t))
	\equiv  \mathcal{C}_{q,N}^{n}\bigl( \{\lambda_i\},\{\sigma_i\}, t \bigr) \, ,\qquad
\label{C def}
\end{eqnarray}
where the $c$-coefficients can contain some integer powers of $\ln(t)$, and $s_*\!=\!1$
for the hypergeometric functions appearing in $F$ structure functions, and $s_*\!=\!0$
for the ones appearing in $G$.
Therefore, replacing  the hypergeometric function by its asymptotic form~(\ref{C def}) captures correctly
the late time limit of the integral~(\ref{J intermediate}) and we can write,
\begin{equation}
\mathcal{J}_{q,N}^{n} \approx a^{-2n} \!
	\int\limits_{1}^{a} \! dt\,
	e^{\frac{iK(t-1)}{a}} t^{-2-N} (t\!-\!1)^{3+2N+2n} \,
	 \mathcal{C}_{q,N}^{n}\bigl( \{\lambda_i\},\{\sigma_i\}, t \bigr)
	+ \mathcal{O}(a^{X_i})  \, ,
\end{equation}
where $\mathcal{C}$ is defined in (\ref{C def})
and $X_{i=F}=0$  and $X_{i=G}=1$ (for notational simplicity we do not expressly include logarithmic corrections
in the order of the estimate).
The resulting integrals can
all be performed. But before doing that we find it far more convenient to switch back to the
integration variable $\tau = t/a$ and switch the order of integration and summation over $n$ in~(\ref{I intermediate}),
\begin{eqnarray}
\mathcal{I}_{q,N} \!\!&\approx&\!\!
	2(-1)^N (N!) a^{2+N}\! \int\limits_{1/a}^{1} \! d\tau\, e^{iK\Delta\tau}
		\tau^{-2-N} (\Delta\tau)^{3+2N}
\label{I intermediate:B} \\
\!\!&\times &\!\!	
	 \sum_{n=0}^{\infty} \frac{(-1)^n(K\Delta\tau)^{2n}}{(2n\!+\!1)!}
	\frac{\Gamma\bigl( n\!+\! \frac{3}{2} \bigr)}
		{\Gamma\bigl( n\!+\!N\!+\frac{5}{2} \bigr)} \,
	\mathcal{C}_{q,N}^{n}\bigl( \{\lambda_i\},\{\sigma_i\}, a\tau\bigr)
+ \mathcal{O}(a^{X_i})
\,.
\quad
\nonumber
\end{eqnarray}
Now Eqs. (16.11.2--16.11.6) of Ref.~\cite{DLMF} allow us to express
the relevant coefficients of asymptotic expansions of the hypergeometric functions
(below we define $X(t)=(t-1)^2/t$) as follows,
\begin{eqnarray}
& &	\hspace{-1.4cm}
\lefteqn{
\mathcal{C}_{1,0}^n
	\Bigl( \bigl\{ \tfrac{7}{2}\!+\!b_N, \tfrac{7}{2}\!-\!b_N \bigr\},
		\bigl\{ 6 \bigr\}, t \Bigr)
}
\\
& &	\hspace{0.98cm}	
= \frac{10(2n\!+\!3)}{\beta} \biggl\{ \frac{1}{X(t)}
	\!+\! \biggl[ 1 \!-\! \frac{4(2n\!+\!1)}{(\beta\!-\!4)} \biggr] \frac{2}{tX(t)}
	\biggr\} \, ,
\nonumber\\
& &	\hspace{-1.4cm}
\lefteqn{
\mathcal{C}_{2,0}^n
	\Bigl( \bigl\{ \tfrac{7}{2}\!+\!b_N, \tfrac{7}{2}\!-\!b_N, 1 \bigr\},
		\bigl\{ 6, 2 \bigr\}, t \Bigr)
=  \frac{10(2n\!+\!3)}{\beta} \biggl\{ \frac{\ln[X(t)]}{X(t)}
}
\nonumber \\
& &	\hspace{0.98cm}
	  + \,\Xi\bigl( \tfrac{7}{2}\!+\!b_N, \tfrac{7}{2}\!-\!b_N, 6,
			n\!+\!\tfrac{5}{2} \bigr) \frac{1}{X(t)}
		\!+\! \frac{8(2n\!+\!1)}{(\beta\!-\!4)} \frac{1}{t X(t)}
	\biggr\} \, ,
\\
& &	\hspace{-1.4cm}
\lefteqn{
\mathcal{C}_{3,1}^n
	\Bigl( \bigl\{ \tfrac{9}{2}\!+\!b_N, \tfrac{9}{2}\!-\!b_N, 1, 1 \bigr\},
		\bigl\{ 7, 2, 2 \bigr\}, t \Bigr)
}
\nonumber \\
& &	\hspace{0.98cm}
	= \frac{12(2n\!+\!5)}{(\beta\!+\!6)} \biggl\{
		\frac{\ln[X(t)]}{X(t)} 
                 \!+\! \Xi\bigl( \tfrac{9}{2}\!+\!b_N, \tfrac{9}{2}\!-\!b_N, 7,n\!+\!\tfrac{7}{2} \bigr)\frac{1}{X(t)}
\nonumber \\
& &	\hspace{0.98cm}
		+ \frac{10(2n\!+\!3)}{\beta} \frac{1}{X(t)^2} \!-\! \frac{40(2n\!+\!3)(2n\!+\!1)}{\beta(\beta\!-\!4)} \frac{1}{t X(t)^2}
	\biggr\} \, ,\qquad
\\
\nonumber \\
& &	\hspace{-1.4cm}
\lefteqn{
\mathcal{C}_{1,1}^n
	\Bigl( \bigl\{ \tfrac{7}{2}\!+\!b_N, \tfrac{7}{2}\!-\!b_N \bigr\},
		\bigl\{ 6 \bigr\}, t \Bigr)
	=  \frac{80(2n\!+\!5)(2n\!+\!3)}{\beta(\beta\!-\!4)} \frac{1}{X(t)^2} \, ,
}
\\
& &	\hspace{-1.4cm}
\lefteqn{
\mathcal{C}_{2,1}^n
	\Bigl(\! \bigl\{ \tfrac{7}{2}\!+\!b_N,\! \tfrac{7}{2}\!-\!b_N, \!1 \bigr\},
		\bigl\{ 6, 2 \bigr\}, t \Bigr)
	\!=\!  \frac{10(2n\!+\!5)}{\beta} \biggl\{\! \frac{1}{X(t)}
		\!-\! \frac{8(2n\!+\!3)}{(\beta\!-\!4)} \frac{1}{X(t)^2}
	\biggr\}   ,
}
\\
& &	\hspace{-1.4cm}
\lefteqn{
\mathcal{C}_{3,1}^n
	\Bigl( \bigl\{ \tfrac{7}{2}\!+\!b_N, \tfrac{7}{2}\!-\!b_N, 1, 1 \bigr\},
		\bigl\{ 6, 2, 2 \bigr\}, t \Bigr)
}
\nonumber \\
& &	\hspace{0.98cm}
	= \frac{10(2n\!+\!5)}{\beta} \biggl\{ \frac{\ln[X(t)]}{X(t)}
	\!+\! \Xi\bigl( \tfrac{7}{2}\!+\!b_N, \tfrac{7}{2}\!-\!b_N, 6,n\!+\!\tfrac{7}{2} \bigr)\frac{1}{X(t)}
\nonumber \\
& &	\hspace{3.4cm}
	+\frac{8(2n\!+\!3)}{(\beta\!-\!4)} \frac{1}{X(t)^2} \biggr\} \, ,
\\
& &	\hspace{-1.4cm}
\lefteqn{
 \mathcal{C}_{3,2}^n
	\Bigl( \bigl\{ \tfrac{9}{2}\!+\!b_N, \tfrac{9}{2}\!-\!b_N, 1, 1 \bigr\},
		\bigl\{ 7, 2, 2 \bigr\}, t \Bigr)
}
\nonumber \\
& &	\hspace{0.98cm}
	= \frac{6(2n\!+\!7)}{(\beta\!+\!6)} \biggl\{
	\frac{\ln[X(t)]}{X(t)}
	\!+\! \biggl[ 1\!+\! \Xi\bigl( \tfrac{9}{2}\!+\!b_N, \tfrac{9}{2}\!-\!b_N,	
			 7, n\!+\!\tfrac{9}{2} \bigr) \biggr] \frac{1}{X(t)}
\nonumber \\
& &	\hspace{3.4cm}
		+ \frac{40(2n\!+\!5)(2n\!+\!3)}{\beta(\beta\!-\!4)} \frac{1}{X(t)^3}
	\biggr\} \, .
\end{eqnarray}
When these expressions are inserted into Eq.~(\ref{I intermediate:B}), the corresponding series over
$n$ can be performed. For example, for $\mathcal{I}_{1,0}$ one obtains,
%
%
%
\begin{eqnarray}
\mathcal{I}_{1,0} \!\!&\approx&\!\!
	\frac{40a}{\beta K}\! \int\limits_{1/a}^{1} \! \frac{d\tau}{\tau^2}\, e^{iK\Delta\tau}
\bigg[
	\Big(\tau+\tfrac{2}{a}\Big)\sin(K\Delta \tau)
-\,\frac{8(K\Delta \tau)}{a(\beta\!-\!4)}\cos(K\Delta \tau)
\bigg]
.\qquad\;
\label{I intermediate:B10}
\end{eqnarray}
This integral can be evaluated and expressed in terms of elementary functions and $\mathcal{G}$.
When expanded in powers of $1/a$, one obtains Eq.~(\ref{I 1 0 : final asymptotic}).

An analogous procedure can be utilized to evaluate the other ${\cal I}_{q,N}$ integrals, yielding
the remaining integrals~(\ref{I 2 0 : final asymptotic}--\ref{I 3 2 : final asymptotic}). 
We do not present here the details of that evaluation.


\section*{Appendix B: The integrals over elementary functions}
\label{app: Integrals over elementary functions}

In this appendix we present some basic steps of the evaluation of the 20 simpler integrals~(\ref{I1def}--\ref{I20def}),
which at late times give~(\ref{I1_asymptotic}--\ref{I20_asymptotic}). The first four integrals are very simple, and
we do not discuss them here. The remaining integrals~(\ref{I5def}--\ref{I20def}) have the following general form 
({\it cf.} Eq.~(\ref{I qN def})), 
\begin{eqnarray}
\mathcal{J}_{\vec n}\Big(\tfrac{k}{H},a\Big)
 \!\!&=&\!\! \frac{a^{n_1}[\ln(a)]^{n_2}H^4}{\pi} \! \int\! d^4x'\,
	e^{ik\Delta\eta - i\vec{k}\cdot\Delta\vec{x}}
\nonumber\\
\!\!&\times&\!\!\!\!
\frac{\partial^{2n_3}}{H^{2n_3}}\left[
	\theta\bigl( \Delta\eta \!-\! r  \bigr)
	(a')^{n_1} [\ln(a')]^{n_2}(H\Delta x)^{2n_4}
		\Big[\!\ln\Bigl(\tfrac{H^2\Delta x^2}{4} \Bigr)\Big]^{n_5}
	\right] ,\qquad\;
\label{J vec n def}
\end{eqnarray}
where $(\vec n)^T=(n_1,n_2,n_3,n_4,n_5)$ with 
$n_1=-1,0,1,2,3$, $n_2=0,1$, $n_3=0,1,2,3$, $n_4=0,1$ and $n_5=0,1$ and 
\begin{equation}
\Delta x^2 = -(\Delta\eta)^2 + r^2\,,\qquad r =\|\vec x-\vec x^{\,\prime}\|
\,,\qquad
\partial^2 = \eta^{\mu\nu}\partial_\mu\partial_\nu = -\partial_0^2+\partial_i^2
\,.
\nonumber
\end{equation}
The first step is to extract the derivatives in~(\ref{J vec n def}) in front of the integral.
In order to do that, one can use the following equlity,
\begin{eqnarray}
 e^{-ik\cdot\Delta x} \partial^2 \!\!&=&\!\! (\partial^2\!+\!2ik\cdot \partial \!-\! k^2)
\,,\quad k\cdot\Delta x = \eta_{\mu\nu}k^\mu \Delta x^\nu =\! -k^0\Delta \eta \!+\! \vec k\cdot (\vec x \!-\!\vec x^{\,\prime})
\quad
\nonumber\\
k^0 \!\!&=&\!\! \|\vec k\|\,,\quad k^2=\eta_{\mu\nu}k^\mu k^\nu = 0
\,.
\label{commuting derivatives}
\end{eqnarray}
The next step is to integrate over the angles, resulting in,
\begin{eqnarray}
\mathcal{J}_{\vec n}\Big(\tfrac{k}{H},a\Big)
 \!\!&=&\!\!4a^{n_1}[\ln(a)]^{n_2}\frac{(\partial^2+2ik\cdot\partial)^{n_3}}{H^{2n_3-4} }
          \! \int_{-1/H}^{-1/(Ha)}\!\!\!d\eta^\prime e^{ik\Delta\eta}
(a')^{n_1} [\ln(a')]^{n_2}
\;\;
\nonumber\\
\!\!&\times&\!\!
                 \int_0^{\Delta \eta} drr^2\,\frac{\sin(kr)}{kr}
(H^2\Delta x^2)^{n_4}
		\Big[\ln\Bigl(\tfrac{H^2\Delta x^2}{4} \Bigr)\Big]^{n_5}
\, .\quad
\label{J vec n: step 1}
\end{eqnarray}
In what follows we perform a substitution of variables to dimensionless quantities,
\begin{equation}
\tau\equiv -H\eta^\prime = \frac{1}{a^\prime}
\,,\quad \Delta \tau = H\Delta \eta = \tau -\frac{1}{a}
\,,\qquad 
 \rho = \frac{r}{\Delta \eta}
\,,\qquad 
K =\frac{k}{H}
\,,
\label{dimensionless variables:def}
\end{equation}
upon which~(\ref{J vec n: step 1}) becomes,
\begin{eqnarray}
\mathcal{J}_{\vec n}\Big(\tfrac{k}{H},a\Big)
 \!\!&=&\!\!4a^{n_1}(-1)^{n_4}[\ln(a)]^{n_2} \frac{(-\partial_0^2-2ik\partial_0)^{n_3}}{KH^{2n_3} }\!\! \int_{1/a}^{1}\!\!\!d\tau e^{iK\Delta\tau}
\tau^{-n_1} [-\ln(\tau)]^{n_2}
\;
\nonumber\\
\!\!&&\!\!\!\! \hspace{-1cm}
              \times  (\Delta\tau)^{2+2n_4} \!\!\int_0^{1}\! d\rho\rho\,\sin(K\Delta\tau\rho)
(1\!-\!\rho^2)^{n_4}
		\Big[\ln\Bigl(\tfrac{1}{4} \Delta\tau^2(1\!-\!\rho^2)\Bigr)\Big]^{n_5}\, ,\quad
\label{J vec n: step 2}
\end{eqnarray}
where we made use of the fact that the integral depends on time but not on space. 
The $\rho$-integral is doable in all cases ($n_4,n_5=0,1$) appearing in~(\ref{I5def}--\ref{I20def}),
and the result can be expressed in terms of elementary functions and the function ${\cal G}(z)$ 
defined in~(\ref{funGdef}). Indeed, if we define,
\begin{eqnarray}
\mathcal{K}_{(n_4,n_5)}(K\Delta\tau) \,\equiv\,  (K\Delta\tau)^{2} \!\!\int_0^{1}\! d\rho\rho\,\sin(K\Delta\tau\rho)
(1\!-\!\rho^2)^{n_4}
		\Big[\!\ln\bigl(1\!-\!\rho^2\bigr)\Big]^{n_5}
\,,\;\;
\label{K integral: definition}
\end{eqnarray}
we have 
\begin{eqnarray}
\mathcal{K}_{(0,0)}(z)  \!&=&\! \sin(z)-z\cos(z)
\label{K integral: 0 0}\\
\mathcal{K}_{(1,0)}(z) \!&=&\!\frac{2}{z^2}\Big[\big(3\!-\!z^2\big) \sin(z)\!-\!3z\cos(z)\Big]
\quad
\label{K integral: 1 0}\\
\mathcal{K}_{(0,1)}(z) &=&\big[ \sin(z)\!-\!z\cos(z)\big]\big[{\rm ci}(2z)-\gamma_E-\ln(z/2)\big]
\nonumber\\
                &&\hspace{-0.1cm}
                  -\,\big[\cos(z)\!+\!z\sin(z)\big]\Big[{\rm si}(2z)\!+\!\frac{\pi}{2}\Big]+2\sin(z)
\qquad
\nonumber\\
               \!&=&\!\bigg[\frac{i-z}{2}\Big(\mathcal{G}(z)+\ln(4)\Big)+i\bigg]e^{-iz}+{\rm c.c.}
\label{K integral: 0 1}
\end{eqnarray}
\begin{eqnarray}
\mathcal{K}_{(1,1)}(z) \!\!&=&\!\!\!\frac{2}{z^2}\bigg\{
\big[ (3\!-\!z^2)\sin(z)\!-\!3z\cos(z)\big]\big[{\rm ci}(2z)-\gamma_E-\ln(z/2)\big]
\nonumber\\
                &&\hspace{0.5cm}
                  -\,\big[(3\!-\!z^2)\cos(z)\!+\!3z\sin(z)\big]\Big[{\rm si}(2z)\!+\!\frac{\pi}{2}\Big]
\nonumber\\
                &&\hspace{0.5cm}
                       +\,(11\!-\!z^2)\sin(z)\!-\!5z\cos(z)
\bigg\}
\qquad
\label{K integral: 1 1}\\
               \!&=&\!\frac{1}{z^2}\bigg[(3i\!-\!3z\!-\!iz^2)\Big(\mathcal{G}(z)\!+\!\ln(4)\Big)\!+\!(11i\!-\!5z\!-\!iz^2)\bigg]e^{-iz}
                \!+\!{\rm c.c.}
\nonumber
\end{eqnarray}
The remaining $\tau$ integral can also be done exactly for all of the integrals in~(\ref{I5def}--\ref{I20def}),
and the results can be expressed in terms of elementary functions and the special functions $\mathcal{G}(z)$,
$\mathcal{M}(x,z)$ and $\mathcal{V}(x,z)$ defined by the integrals~(\ref{funGdef}--\ref{funVdef}).
Finally, upon taking the late time limit
($a\rightarrow \infty$ and $k/(aH)\ll 1$) of these integrals one obtains the results given in
the main text in Eqs.~(\ref{I5_asymptotic}--\ref{I20_asymptotic}).


\end{document}